\documentclass[12pt]{article}
\usepackage{latexsym,amssymb,amsmath}
\usepackage{epsf}
\newtheorem{theorem}{Theorem}[section]
\newtheorem{examp}{Example}[section]
\newtheorem{examps}{Examples}[section]

\newtheorem{lemma}[theorem]{Lemma}
\newtheorem{remark}{Remark}[section]
\newtheorem{remarks}[remark]{Remarks}
\newtheorem{proposition}[theorem]{Proposition}
\newtheorem{definition}{Definition}[section]

\def\begindef{\begin{definition}}
\def\enddef{\end{definition}}
\def\brem{\begin{remark}\rm}
\def\erem{\end{remark}}
\def\erem{\hfill $\square$ \end{remark}}
\def\brems{\begin{remarks}\rm}
\def\erems{\par\noindent $\square$ \end{remarks}}
\def\bexample{\begin{examp}\rm}
\def\eexample{\par\noindent $\square$ \end{examp}}
\def\bexamples{\begin{examps}\rm}
\def\eexamples{\par\noindent $\square$ \end{examps}}
\def\proof{\begingroup\noindent\bf Proof.\ \endgroup}
\def\endpf{\hfill $\square$\\\medskip\\}

\def\rref#1{(\ref{#1})}
\def\de{\partial}
\def\coker{{\rm coker\,}}

\def\Span{{\rm span\,}}

\newcommand{\CH}{{\cal H}}

\newcommand{\WY}{{\mathsf Y}}
\newcommand{\UW}{{\mathsf U}}
\newcommand{\VW}{{\mathsf V}}

\newcommand{\CS}{{\cal S}}
\newcommand{\CT}{{\cal T}}

\makeatletter
\@addtoreset{equation}{section}

\makeatother

\newcommand{\RR}{{\mathbb R}}
\newcommand{\CC}{{\mathbb C}}

\newcommand{\ZZ}{{\mathbb Z}}

\newcommand{\NN}{{\mathbb N}}

\def\al{\alpha}

\def\la{\lambda}        \def\La{\Lambda}
\def\dlt{\delta}

\newcommand{\cmp}[3]{Comm. Math. Phys. {\bf #1} (#2), #3}

\newcommand{\np}[3]{Nucl. Phys. {\bf B #1} (#2), #3}

\newcommand{\lmp}[3]{Lett. Math. Phys. {\bf #1} (#2), #3}

\newcommand{\lanl}[1]{LANL preprint, hep-th/#1}
\newcommand{\jmp}[3]{Jour. Math. Phys. {\bf #1} (#2), #3}

\def\dsl{\displaystyle}

\newcommand{\del}{{\partial}}

\def\Fdb{{Fa\`a di Bruno}}

\newcommand{\dpt}[2]{{\displaystyle{\frac{\partial #1}{\partial t_{#2}}}}}
\newcommand{\Ha}[1]{H^{(#1)}}
\newcommand{\SH}[1]{\wid{H}^{(#1)}}
\newcommand{\SY}{{\wid{Y}}}

\newcommand{\GH}[1]{\wid{\cal H}^{(#1)}}
\newcommand{\sh}[1]{\hat{h}^{(#1)}}

\newcommand{\gh}[1]{\hat{\mathfrak h}^{(#1)}}
\newcommand{\ggh}{\hat{\mathfrak h}}
\newcommand{\Bring}{{B_{[x\varphi]}}}

\def\ha#1{h^{(#1)}}

\def\ger{hierarch}

\def\bih{bihamiltonian}

\def\ham{hamiltonian}

\def\ger{hierarch}

\newcommand{\wid}[1]{\widehat{#1}}
\begin{document}
\begin{flushright}
Ref. SISSA 152/1999/FM

nlin.SI/0001052
\end{flushright}
\vspace{1.truecm}
\begin{center}
{\huge Super KP equations and Darboux transformations:
another perspective on the Jacobian Super KP \ger y}
\end{center}
\vspace{1,2truecm}
\makeatletter
\begin{center}
{\large
Gregorio Falqui, Cesare Reina
and Alessandro Zampa\\ \bigskip
SISSA, Via Beirut 2/4, I-34014 Trieste, Italy}\\
E--mail: falqui@sissa.it,
reina@sissa.it,
zampa@fm.sissa.it\\
\vspace{.3truecm}
\end{center}
\makeatother
\vspace{1.5truecm}
\noindent {\bf Abstract}. 
We generalize to the supersymmetric case the representation of the KP \ger y as
a set of conservation laws for the generating series of the conserved
densities. We show that the hierarchy so obtained is isomorphic to
 the JSKP of Mulase and Rabin. 
We identify its ``bosonic content'' with the so--called
Darboux--KP \ger y, which geometrically encompasses the theory of
Darboux--B\"acklund transformations, and is an extension both of the KP theory
and of the modified KP theory. Finally, we show how the
\ger y can be linearized and how the supersymmetric counterpart of a
wide class of rational solution can be quite explicitly worked out. 
\vspace{.1truecm}
\section{Introduction}\label{setc:0}
Supersymmetric extensions of integrable \ger ies of PDE's are by now a
well-studied subject (see,
e.g.,~\cite{KupS,MaRad,Mu,Rab,Mat88,LeC89,IbMaMe96,LiMa}).
 In particular, for the KP equations, two different extensions have
 emerged: the {\em Manin--Radul} SKP~\cite{MaRad} (MRSKP), 
and the {\em Jacobian} SKP of Mulase and
Rabin~\cite{Mu,MuRab} (JSKP). The most remarkable and best know differences
between the two \ger ies are the fact that the flows of MRSKP are a
representation of a super Heisenberg algebra, while those of JSKP are
supercommuting ones, and the behavior of their algebro--geometric solution.
Indeed, the JSKP describes linear flows on the (super) Jacobian manifold of
a super algebraic curve $\wid C$, while the MRSKP flows involve, in general, 
a motion on the space of moduli of $\wid C$. Actually, the latter is perhaps 
the strongest
motivation that led Mulase and Rabin to modify the (previously discovered)  
Manin--Radul supersymmetric extension of the KP \ger y. \\
From the point of view of the present paper, it is more relevant another
difference between the two approaches: the Manin--Radul theory  
concerns the supersymmetric extension of the {\em Lax representation}:
\begin{equation}\label{eq:1.1}
\dpt{L}{k}=\big[\big(L^k\big)_+,L\big]
\end{equation}
of the KP \ger y on the space of pseudo-differential operators 
in one dimension.
In the approach of Mulase and Rabin, one starts instead from
the {\em Sato representation} of the KP equations on the Volterra group of
Dressing operators,
\begin{equation}\label{eq:1.2}
\dpt{S}{k}=-\big( S\del_x^k S^{-1} \big)_- S\>,
\end{equation}
where $S$ and $L$ are related by the well--know dressing formula
\begin{equation}\label{eq:1.3}
L=S\del_xS^{-1}\>.
\end{equation}
This paper is based on the representation of the KP theory as a set of
conservation laws~\cite{Wil81}, which has been recently studied~\cite{FMP}
as an outgrowth of the application of the Gel'fand--Zakharevich
theorem~\cite{GZ} to infinite dimensional integrable systems. Namely,
the KP equations are written as the conservation laws
\begin{equation}\label{eq:1.4}
\dpt{h}{k}=\del_x \Ha{k}\>,
\end{equation}
where
\begin{equation}\label{eq:1.4bis}
h=z+\sum_{l=1}^{\infty} \frac{h_l}{z^l}
\end{equation}
is the generating series of the conserved densities of the KP theory, and
$\Ha{k}$ are their current densities. These currents are written as
suitable linear combinations of the \Fdb\ monomials
\begin{equation}\label{eq:1.5}
\ha{k}=\big(\del_x+h\big)^k\cdot 1
\end{equation}
associated with $h$.\\
The supersymmetric extension of the equations~\rref{eq:1.4} will be  written as
\begin{equation}\label{eq:1.6}
\dpt{\hat h}{\al}=(-1)^\al \dlt \SH{\al}\>,
\end{equation}
and called  Hamiltonian Super KP \ger y (HSKP).
Here $\dlt=\del_\varphi+\varphi\del_x$ 
is the square root on the super-circle
$S^{1|1}$ of the $x$--derivative $\del_x$, well known from string and
superconformal field theories (see, e.g., \cite{Frie}), the odd superfield  
$\hat h$ replaces the generator $h$ of the local \ham\ densities, and
$\SH{\al}$  are suitable supersymmetric extensions of the currents $\Ha{j}$. 
In the first part of the paper we show that, as it happens in the
ordinary case, equations~\rref{eq:1.6} admit a representation--extension to a
family of dynamical systems with $\NN^2\times \NN^2$ variables 
(of which half are even and half are odd), to be called the 
{\em Super Central System} (SCS).
It is a counterpart of the Sato system on the infinite--dimensional
Grassmann manifold~\cite{SS,DJKM,SW,MuAlg,Tks}, and the Super KP 
equations~\rref{eq:1.6} 
can be considered as a kind of ``reduction'' of SCS. 
Then we show how the \ger y~\rref{eq:1.6} can be identified, by means of a non 
trivial coordinate change, with the Jacobian Super KP \ger y.
 
This is however only the first topic we want to discuss in this paper. 
The second one concerns the relation of the HSKP \ger y with the
theory of Darboux transformations. The fact that supersymmetry has an 
intriguing relation with the theory of 
Darboux transformations is a well established one. For instance, the 
classical problem of factorizing the Schr\"odinger operator $-\del_x^2+u(x)$
into first order factors gives rise to a super algebra (see~\cite{MS}, \S2,
and references quoted therein). Another signal of this fact comes from
Mulase's paper~\cite{SolvBirk}, where it was shown how the modified KP
equation can be obtained from the (Manin--Radul) SKP by means of a process of
elimination of odd variables. 

We shall show that HSKP provides a natural
framework to discuss such issues. We shall use the
geometrical setting of~\cite{MPZ}, where  the classical subject
of Darboux--B\"acklund transformations and Miura--like maps is approached in
a rather unconventional way that can be summarized as follows. 
Instead of searching directly for a symmetry of an evolutionary equation $X$
defined on a manifold $M$,  one tries to find a {\em covering}, that is
another evolutionary equation, defined on a bigger manifold
$N$, related to $X$ by two maps $\pi,\sigma: N \to M$, 
such that
\begin{equation}
X=\pi_*Y=\sigma_*Y\>. 
\end{equation}  
In~\cite{MPZ} a
covering for the KP \ger y, called Darboux--KP (DKP) \ger y, was constructed
as  a \ger y defined on the phase space of pairs of Laurent series $(h,a)$
with suitable asymptotics and the
(generalized) Miura transformation was defined to be $h\mapsto
\sigma(a,h)=h+{a_x}/{a}.$
Another application of this formalism has been used
in~\cite{FMP} to linearize 
the equations (the so called {\em Central System (CS)})
induced by the flows~\rref{eq:1.4} on the currents
$\Ha{j}$. This method  has been exploited (see~\cite{fmpz}) to 
construct a wide class of  solutions of KP admitting a polynomial
$\tau$--function.

In this paper we will consider the supersymmetric counterpart
of the geometric theory of the  Darboux transformations. 
Firstly, we will show that the DKP system is actually the bosonic content 
of the even flows of HSKP (and hence of JSKP). Then we will construct 
a Darboux covering for HSKP and related ``rational'' reductions.  
Finally,  we will use the technique of Darboux covering
to linearize the Super analogue SCS of CS, exploiting this result in a
quite explicit description of a wide class of  non trivial  solutions of HSKP
of rational type. 

The detailed plan of the paper is the following: in  Section~\ref{sect:1},
after having briefly recalled the basis of the (bi)hamiltonian set up for the
KP theory, we will introduce the phase space for HSKP and define the \ger y.
In Section~\ref{SCS} we will discuss the fundamental properties of HSKP, and
we will introduce the Super Central System SCS as the
dynamical system obeyed by the currents of the theory when the \Fdb\ generator
$\hat{h}$ evolves along HSKP. In Section~\ref{sect:scs-hskp} we will show how
solutions of HSKP can be obtained from solutions of SCS, and
in Section~\ref{others} we will briefly discuss
how HSKP can be seen as a particular form of the Jacobian SKP \ger y of Mulase 
and Rabin, by comparing  the wave functions associated with the two theories.
In Section~\ref{sect:1-1}, we will show how a super extension of the KdV
equation can be obtained as a suitable reduction of HSKP; this will give us a
concrete clue to the rest of the paper. Indeed,
from Section~\ref{Darboux} onwards we will turn our attention to the 
method of Darboux coverings. We will first recall the
setting of the ordinary bosonic case, and then identify the bosonic part of
HSKP with the DKP \ger y of~\cite{MPZ}. We will also point out the specific 
form of the generalized Miura transformation. Furthermore, we will construct a
Darboux  covering of HSKP, and briefly discuss some reductions of the 
latter. Finally, in Section~\ref{linear} we will show how the 
equations can be explicitly linearized, and discuss a specific class of
solutions depending rationally on a finite number of times.
\section{The GZ approach to KP and its 
supersymmetric extension}\label{sect:1}
The technique that plays a prominent role in the bi-Hamiltonian approach
to KP is the Gelfan'd Zakharevich 
method of Poisson pencils to construct integrable Hamiltonian
systems \cite{GZ}. In such a scheme one considers a 
manifold $M$ endowed with a pencil $P_\la=P_1-\la P_0$
of Poisson structures,
and studies the {\em Casimir functions} of the pencil.
Such a Casimir function $H_\lambda$ 
is a (non-constant) function on $M$, which depends also 
on the parameter $\lambda$, such that $P_\lambda dH_\lambda=0$ 
for every value of $\lambda$.
When $M$ is an $2n+1$--dimensional  manifold endowed with a Poisson pencil of
maximal rank, $P_\lambda$ has a unique Casimir function $H_\lambda$, 
which is a polynomial in $\lambda$ of degree $n$,
$$H_\lambda=H_0\lambda^n+H_1\lambda^{n-1}+\cdots+H_n.$$
Its leading coefficient $H_0$ is the Casimir of $P_0$,
while the ``constant term'' $H_n$ is the Casimir of $P_1$.
The coefficients $H_j$ satisfy the recurrence relations
$$P_1dH_{j+1}=P_0dH_j$$
and therefore are in involution with respect to all the brackets of the pencil.

In the realm of infinite dimensional systems, 
the KdV theory is perhaps the best known prototype  
of a GZ hierarchy \cite{CMP,GZ-CPAM}.
Here, the manifold $M$ is the space of $C^\infty$ functions
on the circle $S^1$, 
and the Poisson pencil, given as a one-parameter family of skew-symmetric
maps from the cotangent to the tangent bundle, reads
$$\dot{u}=(P_\lambda)_uv=-{\frac12}v_{xxx}+2(u+\lambda)v_x+u_xv,$$
where $x$ is a coordinate on $S^1$, $u$ represents a point of $M$, and $\dot
u$ and $v$ are respectively a vector and a 
a covector at $u.$
It turns out \cite{FMP} that if $h$ and $v$ are
series in $z=\sqrt{\lambda}$ of the form
\begin{equation}\label{eq:hv}
h(z)=z+\sum_{j>0}h_jz^{-j}, \quad
v=1+\sum_{l>0}v_l/z^{2l}
\end{equation} 
that provide the unique solutions of
the Riccati system
\[
\left\{
\begin{split}
&h_x+h^2=u+z^2
\\
&-{{1}/{2}}v_x+hv=z\end{split}\right. ,
\]
then $v(z)$ is the series
representing the differential of the Casimir function of the 
Poisson pencil of KdV, 
which, in turn,  is given by the integral 
\begin{equation}\label{eq:casint}
H_\la =2z\int_{S^1}hdx.
\end{equation}
The GZ hierarchy associated with $H_\la$ on $M$ admits several representations.
The one we are interested in can be expressed by saying that the local
Hamiltonian density $h(z)$ must obey local conservation laws of the form
$$\dpt{}{j} h=\del_xH^{(j)},$$
where the ``current densities'' $H^{(j)}$ are given by
\begin{equation}\label{eq:Hkdv}
H^{(2j)}=\lambda^j\>\mbox{ and  }\quad
H^{(2j+1)}=-\frac{1}{2}(\lambda^jv)_{+,x}+h(\lambda^jv)_+,
\end{equation}
the subscript $+$ meaning to take the positive part of the expansion in
powers of $z$. 
Equation~\rref{eq:Hkdv} can also be written as 
$$H^{(2j+1)}=
z^{2j}\left(-\frac{1}{2}v_x+hv\right)+\frac{1}{2}(z^{2j}v)_{-,x}-
h(z^{2j}v)_-$$
or
$$H^{(2j+1)}=\sum_{l=1}^j\left[-\frac{1}{2}v_{j-l,x}(z^{2l}\cdot 1)+
v_{j-l}(z^{2l}\cdot h)\right].$$
The first of these two expressions shows that $H^{(j)}=z^j+O(z^{-1})$, since by
the second Riccati equation above we have
$z^{2j}\left(-\frac{1}{2}v_x+hv\right)=z^{2j+1}.$ 
The interpretation of the second entails that
the currents $\Ha{j}$ are the unique combinations
$$H^{(j)}=\sum_{k=0}^jc^j_kh^{(k)}$$ 
of the \Fdb\ iterates
$\ha{j}$ of $h^{(0)}=1$ at $h$, defined by
\begin{equation}\label{eq:faa}
h^{(j+1)}=(\del_x+h)h^{(j)},
\end{equation}
that admit the asymptotic expansion 
$H^{(j)}=z^j+O(z^{-1})$.\\
The relevance of this result is that
the currents $H^{(j)}$ can be constructed without 
requiring that $h$ is a solution of the
Riccati equation. Therefore one can define the KP \ger y as follows.
\begindef
Let $h$ be a monic (formal) Laurent series in $z^{-1}$
$$h:=z+\sum_{j>0}h_jz^{-j},$$
whose coefficients $h_j$ belong to $C^\infty(S^1)$, and 
consider its \Fdb\ iterates $h^{(j)}$.
Denote by $W$ the span over $C^\infty(S^1)$ of the order Fa\`a di
Bruno iterates (or monomials) 
$$W:=\Span_{C^\infty(S^1)}\{h^{(k)},\ k\ge 0\}$$
and introduce the ``current densities'' $H^{(k)}$ by requiring them to be
the unique elements of $W$ of the form
$$H^{(k)}=z^k+\sum_{j>0}H^k_jz^{-j}.$$
The {\em KP hierarchy} is defined to be the set of conservation laws
\begin{equation}
  \label{eq:KPdef}
\dpt{}{k} h =\de_xH^{(k)}.  
\end{equation}
\enddef
We observe that $H^{(1)}=h$, so we can identify the first time $t_1$ with
$x$. Moreover, it can be proven that this definition is completely equivalent
to the one given in the framework of pseudo-differential operators (see,
e. g.,~\cite{D} and references quoted therein).

We end this  review of the \bih\ set-up of the KP \ger y with the
notion of the {\em Central System} (CS).  The operators
$\del_{t_k}+\Ha{k}$ satisfy the invariance condition
\[
\big(\del_{t_k}+\Ha{k}\big)W\subset W
\]
and the commutativity property $\big[
\del_{t_k}+\Ha{k},\de_{t_j}+\Ha{j}\big]=0$. This entails that along the KP
flows~\rref{eq:KPdef} the currents $\Ha{k}$ satisfy the following evolutionary 
equation:
\begin{equation}
  \label{eq:CSx}
  \dpt{}{j}\Ha{k}+\Ha{j}\Ha{k}=\Ha{j+k}+\sum_{l=1}^k
  H^j_l\Ha{k-l}+\sum_{l=1}^{j} H^k_l\Ha{j-l}\>.
\end{equation}
\begindef
Let $\CH$ be the space of sequences $\{\Ha{0},\Ha{1},\Ha{2},\Ha{3},\ldots\}$ 
where $\Ha{0}=1$ and the $\Ha{j}$'s are of the form:
\[
\Ha{j}=z^j+\sum_{l\ge 1} \frac{H^j_l}{z^l},\> j\ge 1
\]
The Central System is the \ger y of dynamical system on $\CH$
defined by equation~\rref{eq:CSx}. It turns out that the vector fields of CS
commute among themselves~\cite{CS}.
\enddef
 In the next  Section we define the extension to the ($N=1$) 
supersymmetric case
of the constructions herewith outlined. This will lead us to the definition
of a supersymmetric extension of the KP \ger y.  
We will refer to such a SKP theory 
as  Hamiltonian Super KP, to  keep track of its (albeit remote) \ham\ origins. 
Later (see Section~\ref{others})   we will show how to  identify 
our HSKP with the Jacobian
super KP hierarchy of Mulase and Rabin \cite{Mu,Rab,BerRab,Tak}) . \\
\subsection{The definition of the Hamiltonian super KP hierarchy}\label{FdBSKP}
Let us start by fixing  some notations (see, e.g.,~\cite{Gauge} for more
details on supergeometry), to be used throughout the paper.
We denote  by $\Lambda$ a generic Grassmann algebra over $\mathbb C$.
This is required by functorial properties of supersymmetry~\cite{Schw},
but in this work it will play a spectator role, and can be thought of as a
fixed algebra. 
We supplement the bosonic spectral parameter $z$ with its fermionic
``super-partner'' $\theta$, and replace the circle by its super 
analog $S^{1\vert 1}$ endowed with coordinates
$x$, $\varphi$. To simplify notations we call $B_{[x\varphi]}$ 
the ring $C^\infty(S^{1\vert 1},\Lambda)$
of smooth functions on $S^{1\vert 1}$ with values in $\Lambda$ (or a suitable
``version'' of it, like the space of $\La$-valued
functions on $\RR^{1|1}$ vanishing at
infinity, or even the space $\CC[[x,\varphi]]\otimes \La$
of formal series in $x,\varphi$.) 
Finally, we denote by $\bar{f}$ the parity of a
homogeneous element $f$, e.g. $\bar{z}=\bar{x}=0$,
$\bar{\theta}=\bar{\varphi}=1$.

The Hamiltonian super KP hierarchy is defined in terms of the super
Fa\`a di Bruno generator $\hat{h}$  and the odd derivation operator 
$\delta:=\de_\varphi+\varphi\de_x$ taking the place of $\de_x.$
The generator $\hat{h}$ is an odd formal 
Laurent series in $z^{-1}$ and $\theta$ with
coefficients in $B_{[x\varphi]}$ of the form
\begin{equation}\label{eq:hath}
\hat{h}(z,\theta;x,\varphi):=
\nu(z;x)+\theta a(z;x)+\varphi h(z;x)+(\theta\varphi)\psi(z;x),
\end{equation} 
We specify exactly the content of the components $\nu$, $a$, $h$ and $\psi$,
by analogy with the KP formalism, requiring that:
\begin{enumerate}
\item All equations are homogeneous with respect to the grading specified by
\[
[\theta]=\frac12,\> [z]=1,\> [\varphi]=-\frac12,\> [x]=-1,\> [\hat
h]=[\delta]=\frac12,\>[t_{j}]=\frac{j}{2},
\]
and no field of negative weight enters the theory. 
\item It is possible to identify the second time $t_2$ with $x$.\footnote{
The relation of $t_1$ with $\varphi$ will be  however much subtler. We will
discuss it in Section~\ref{sect:scs-hskp}.}
\item There exist suitable ``super current densities'' $\SH{k}$, with
asymptotic behavior
\[
\SH{2j+p}\sim z^j\theta^p+O(1/z), \> j\in \NN, p\in\{0,1\}.
\]
\end{enumerate}
It turns out that these requirements can be satisfied if 
the following simple and convenient choices are made:
\begin{description}
\item[i)] $a$
is holomorphic in $z^{-1}$ with constant and invertible 
zeroth order coefficient (which we assume to be equal to $1$):
$a(z;x):=1+\sum_{j>0}a_j(x)z^{-j}.$
\item[ii)] 
$h$ has the usual form $h(z;x):=z+\sum_{j>0}h_j(x)z^{-j}.$
\item[iii)] $\nu$ and $\psi$ are of the form
$$\nu(z;x):=\sum_{j>0}\nu_j(x)z^{-j}, 
\qquad \psi(z;x):=\sum_{j>0}\psi_j(x)z^{-j}.$$
\end{description}
With the super Fa\`a di Bruno generator $\hat{h}$ we associate (for
$k\in{\mathbb N}$) its iterates
$$\left\{\begin{array}{l}
\hat{h}^{(k+1)}:=(\delta+\hat{h})\cdot\hat{h}^{(k)} \\
\hat{h}^{(0)}:=1
\end{array}\right..$$
The following lemma shows that the even Fa\`a di Bruno iterates, apart  from
their nilpotent components, are essentially  
the usual Fa\`a di Bruno monomials.
\begin{lemma}\label{lem21}
Let
$\hat{f}:=\delta(\hat{h})=
h-\theta\psi+\varphi\nu_x-(\theta\varphi)a_x$.
Then, for any $k\in{\mathbb N}$
\begin{equation}\label{eq:sfdbrecrel}
\left\{\begin{array}{l}
\hat{h}^{(2k+2)}=(\de_x+\hat{f})\cdot\hat{h}^{(2k)}=
(\de_x+\hat{f})^{k+1}\cdot 1 \\
\hat{h}^{(2k+3)}=(\de_x+\hat{f})\cdot\hat{h}^{(2k+1)}=
(\de_x+\hat{f})^{k+1}\cdot\hat{h}
\end{array}\right..
\end{equation}
\end{lemma}
\proof
We have
$$(\delta+\hat{h})^2=
\delta^2+\delta(\hat{h})-\hat{h}\delta+\hat{h}\delta+\hat{h}^2=
\delta^2+\delta(\hat{h})=\de_x+\hat{f},$$
so we get
$$\hat{h}^{(2k+2)}=(\delta+\hat{h})^2\cdot\hat{h}^{(2k)}=
(\de_x+\hat{f})\cdot\hat{h}^{(2k)}=
(\delta+\hat{h})^{2k+2}\cdot 1=(\de_x+\hat{f})^{k+1}\cdot 1$$
 and
$$\hat{h}^{(2k+3)}=(\delta+\hat{h})^2\cdot\hat{h}^{(2k+1)}=
(\de_x+\hat{f})\cdot\hat{h}^{(2k+1)}=
(\delta+\hat{h})^{2k+2}\cdot\hat{h}=(\de_x+\hat{f})^{k+1}\cdot\hat{h}.$$
\endpf
For later use we express the Fa\`a di Bruno iterates as
\begin{equation}\label{eq:formhh}
\left\{\begin{array}{l}
\hat{h}^{(2k)}=
h^{(k)}-\theta\psi^{(k)}+\varphi\omega^{(k)}-(\theta\varphi)b^{(k)} \\
\hat{h}^{(2k-1)}=
\nu^{(k)}+\theta a^{(k)}+\varphi d^{(k)}+(\theta\varphi)\chi^{(k)}
\end{array}\right.,
\end{equation}
where the components are Laurent series of the form
$$\begin{array}{ll}
\nu^{(k)}=\sum_{j>0}\nu^{(k)}_jz^{k-j-1}, &
h^{(k)}=z^k+\sum_{j>0}h^{(k)}_jz^{k-j-1} \\
a^{(k)}=z^{k-1}+\sum_{j>0}a^{(k)}_jz^{k-j-1}, &
\psi^{(k)}=\sum_{j>0}\psi^{(k)}_jz^{k-j-1} \\
d^{(k)}=z^k+\sum_{j>0}d^{(k)}_jz^{k-j-1}, &
\omega^{(k)}=\sum_{j>0}\omega^{(k)}_jz^{k-j-1} \\
\chi^{(k)}=\sum_{j>0}\chi^{(k)}_jz^{k-j-1}, &
b^{(k)}=\sum_{j>0}b^{(k)}_jz^{k-j-1}
\end{array}$$
and can be computed by recurrence according to the rules displayed in Table 1
\begin{center}
\begin{table}[hbt]
\begin{tabular}{|| ll ||}
\hline\hline
 & \\
\qquad\qquad Table 1&\\
 & \\
\hline\hline
 & \\
$\left\{\begin{array}{l}
\nu^{(k+1)}=(\de_x+h)\nu^{(k)} \\
\nu^{(1)}=\nu
\end{array}\right.$ &$
\left\{\begin{array}{l}
h^{(k+1)}=(\de_x+h)h^{(k)} \\
h^{(0)}=1
\end{array}\right.$ 
\\ & \\
$\left\{\begin{array}{l}
a^{(k+1)}=(\de_x+h)a^{(k)}-\psi\nu^{(k)} \\
a^{(1)}=a
\end{array}\right.$  &
$\left\{\begin{array}{l}
\psi^{(k+1)}=(\de_x+h)\psi^{(k)}+\psi h^{(k)} \\
\psi^{(0)}=0
\end{array}\right.$ \\& \\
$\left\{\begin{array}{l}
 d ^{(k+1)}=(\de_x+h) d ^{(k)}+\nu_x\nu^{(k)} \\
 d ^{(1)}=h
\end{array}\right.$ &
$\left\{\begin{array}{l}
\omega^{(k+1)}=(\de_x+h)\omega^{(k)}+\nu_xh^{(k)} \\
\omega^{(0)}=0
\end{array}\right.$ \\& \\
$\left\{\begin{array}{l}
\chi^{(k+1)}=(\de_x+h)\chi^{(k)}+\psi d ^{(k)}\\
\phantom{\chi^{(k+1)}=}+\nu_x a^{(k)}-a_x\nu^{(k)} \\
\chi^{(1)}=\psi
\end{array}\right.$ &
$\left\{\begin{array}{l}
b^{(k+1)}=
(\de_x+h)b^{(k)}-\psi\omega^{(k)}\\
\phantom{b^{(k+1)}=} -\nu_x\psi^{(k)}+a_xh^{(k)}\\
b^{(0)}=0
\end{array}\right.$ \\ &\\
\hline\hline
\end{tabular}
\end{table}
\end{center}

By analogy with the KP hierarchy, we introduce the space
\begin{equation}\label{eq:WB}
W_{B_{[x\varphi]}}:=\Span_{B_{[x\varphi]}}\{\hat{h}^{(k)},\, k\in{\mathbb N}\}
\end{equation}
and prove the existence of the super currents with the desired asymptotic
behavior. 

\begin{proposition}\label{currents}
Let $\hat{h}$ and $W_{B_{[x\varphi]}}$ be defined as 
in equations~\rref{eq:hath} and~\rref{eq:WB}.
There exists a basis $\{\SH{k},\, k\in{\mathbb N}\}$ of 
$W_{B_{[x\varphi]}}$
with
$$\left\{\begin{array}{l}
\SH{2k}=
z^k+\sum_{j>0}\left(\wid{H}^{2k}_{0,j}(x,\varphi)z^{-j}+
\wid{H}^{2k}_{1,j}(x,\varphi)\theta z^{-j}\right) \\
\SH{2k+1}=
\theta z^k+\sum_{j>0}\left(\wid{H}^{2k+1}_{0,j}(x,\varphi)z^{-j}+
\wid{H}^{2k+1}_{1,j}(x,\varphi)\theta z^{-j}\right)
\end{array}\right..$$
\end{proposition}

\proof
By definition of $W_{B_{[x\varphi]}}$ we see that
\begin{equation}\label{eq:primeH}
\left\{\begin{array}{l}
\SH{0}=1 \\
\SH{1}=\hat{h}^{(1)}-\varphi\hat{h}^{(2)} \\
\SH{2}=\hat{h}^{(2)}
\end{array}\right..
\end{equation}
The others can be computed recursively: suppose we have defined
$\SH{j}$ for $0\le j<k$; if $k=2n$ is even then
\begin{equation}\label{eq:recpari}
\SH{k}=\hat{h}^{(k)}-
\sum_{j=1}^{n-1}\left(h^{(n)}_j+
\varphi\omega^{(n)}_j\right)\SH{k-2j-2}-
\sum_{j=1}^{n-1}\left(\psi^{(n)}_j+
\varphi b^{(n)}_j\right)\SH{k-2j-1},
\end{equation}
while if $k=2n-1$ is odd then
\begin{equation}\label{eq:recdispa} 
\begin{array}{l}\SH{k}=\hat{h}^{(k)}-\varphi\hat{h}^{(k+1)}-
\sum_{j=1}^{n-1}\left(\nu^{(n)}_j+
\varphi(d^{(n)}_j-h^{(n)}_j)\right)\SH{k-2j-1} \\
\phantom{\hat{H}^{(k)}=}-\sum_{j=1}^{n-1}\left(a^{(n)}_j+
\varphi(\chi^{(n)}_j-\psi^{(n)}_j)\right)\SH{k-2j}
\end{array}.
\end{equation}
We have thus prepared all the ``ingredients'' needed for the following
\begindef{\bf (HSKP)}
Let $\hat{h}$ be defined by~\rref{eq:hath}, 
and compute  its Fa\`a di Bruno iterates $\hat{h}^{(k)}$ and the basis
$\{\SH{k},\ k\ge 0\}$ of $W_{B_{[x\varphi]}}$ as explained in 
Proposition~\ref{currents}.
The {\em Hamiltonian super KP hierarchy} is the set of ``super
conservation laws''
\begin{equation}\label{eq:defHSKP}
\de_{t_k}\hat{h}=(-1)^k\delta\SH{k},\qquad k>0\>.
\end{equation}
\enddef
Notice that, according to the last line of~\rref{eq:primeH}, one has
$\SH{2}=\sh{2}=\delta\hat{h}$. Hence the $t_2$ equation of motion of HSKP is
\[
 \de_{t_2}\hat{h}=\delta\SH{2}=\delta(\delta\hat{h})=\del_x\hat{h},
\]
that is, indeed, $t_2$ can be identified with $x$.
\subsection{The super central system}\label{SCS}
The first property to be verified is the compatibility of
the evolution equations~\rref{eq:defHSKP}.
While checking this, we shall find that the hierarchy has some very
useful properties which allow us to describe it as producing flows on the
super universal Grassmannian.
This will be accomplished by the introduction of a dynamical system
tightly connected with HSKP.
Let us first work out some simple consequences
of the definition of the hierarchy.

The evolution equations~\rref{eq:defHSKP} are simply the 
super-commutativity conditions
$$[\delta+\hat{h},\de_{t_k}+\SH{k}]=0$$
and imply that
\begin{equation}\label{eq:S-inv}
\left(\de_{t_k}+\SH{k}\right)\cdot W_{B_{[x\varphi]}}\subset 
W_{B_{[x\varphi]}}.
\end{equation}
Indeed,
\begin{equation*}\begin{split}
\left(\de_{t_k}+\SH{k}\right)&\cdot\hat{h}^{(l)}  =
\left(\de_{t_k}+\SH{k}\right)\cdot\left(\delta+\hat{h}\right)^l\cdot 1 
\\ & =  (-1)^{kl}\left(\delta+\hat{h}\right)^l\cdot
\left(\de_{t_k}+\SH{k}\right)\cdot 1 
= (-1)^{kl}\left(\delta+\hat{h}\right)^l\cdot\SH{k}\end{split}
\end{equation*}
and by definition $\SH{k}\in W_{B_{[x\varphi]}}$,
$(\delta+\hat{h})\cdot W_{B_{[x\varphi]}}\subset W_{B_{[x\varphi]}}$.
In turn, this implies the  ``abelian zero curvature'' equation:
\begin{equation}\label{eq:0cur}
\de_{t_j}\SH{k}=(-1)^{jk}\de_{t_k}\SH{j},
\end{equation}
as the following simple argument shows.

Denote by $V_{B_{[x\varphi]}}$ the space of
Laurent series in $z^{-1}$ and $\theta$ with coefficients in $B_{[x\varphi]}$ 
and by $V_{B_{[x\varphi]}}^-$ its subspace of formal power series without 
``constant term'', i.e. starting from $z^{-1}$ and $\theta z^{-1}$. We 
have the decomposition \begin{equation}\label{eq:deco1}
V_{B_{[x\varphi]}}=W_{B_{[x\varphi]}}\oplus V_{B_{[x\varphi]}}^-.
\end{equation}
Then, by Equation~\rref{eq:S-inv}, $\de_{t_j}\SH{k}$ is the
$V_{B_{[x\varphi]}}^-$-component of  $-\SH{j}\SH{k}$, while
$\de_{t_k}\SH{j}$ is the $V_{B_{[x\varphi]}}^-$-component of
$-\SH{k}\SH{j}=-(-1)^{jk}\SH{j}\SH{k}$, thus proving 
equations~\rref{eq:0cur}. \\
Thanks to this property and the commutativity of the operators $\delta$ and
$\del_{t_j}$,  we obtain the compatibility of the evolution equations
$$\de_{t_j}\de_{t_k}\hat{h}=(-1)^{j+k}\delta\de_{t_j}\SH{k}=
(-1)^{jk+j+k}\delta\de_{t_k}\SH{j}=
(-1)^{jk}\de_{t_k}\de_{t_j}\hat{h}.$$
This result finally entails
the supercommutativity of the operators $\del_{t_k}+\SH{k}$,
$$[\de_{t_j}+\SH{j},\de_{t_k}+\SH{k}]=0.$$

We notice that it is possible to describe the theory
in terms of the super-currents $\wid{H}^{(k)}$'s only, avoiding the  
introduction of the
super-space variables $x$ and $\varphi$ and the super-derivative $\delta$
which up to now played a special role.
It is by doing this that the super universal Grassmannian arises.
Let us first of all recall its definition~\cite{Schw,BerRab,UeYa}.\\
Denote by $V:=\Lambda((z^{-1}))\oplus\Lambda((z^{-1}))\cdot\theta$
the quotient ring of the ring of formal power series in $z^{-1}$ and
$\theta$ over\footnote{To make contact with the 
  previous definitions, notice that 
$V_{B_{[x\varphi]}}=V\otimes_\Lambda \Bring$.} $\Lambda$, 
and let $V_+:=\Lambda[z,\theta]$,
$V_-:=\Lambda[[z^{-1},\theta]]\cdot z^{-1}.$
$V$ has a natural filtration
$$\cdots\subset V_{j-1}\subset V_j\subset V_{j+1}\subset\cdots,$$
where $V_j=z^{j+1}V_-$, which makes it and its $\Lambda$-submodule $V_+$
complete topological spaces.
Then, the {\em super Grassmannian} $SGr_\Lambda:=SGr_\Lambda(V,V_+)$ is the
set of closed free $\Lambda$-submodules $W$ of $V$ which are compatible with
$V_+$ in the sense that the restriction $\pi_W$ of the projection
$\pi_W:V\to V_+$ to $W$ is a Fredholm operator, i.e. its
kernel (respectively cokernel) is a $\Lambda$-submodule (respectively a
$\Lambda$-quotient module) of a finite rank free $\Lambda$-module.
As usual~\cite{Schw,SW}, 
$SGr_\Lambda$ is the disjoint union of the denumerable set of its
components labelled by the index $i_W$ of $\pi_W$.

We exploit our formul\ae\  and the concept of super universal Grassmannian
by means of the following

\begindef{\bf (SCS)}
Let $M$ be the set of sequences $\{\SH{k}\}_{k\ge 0}$ of formal
Laurent series with coefficients in $\Lambda$ admitting the following
expansion in $z$:
$$\left\{\begin{array}{l}
\SH{2k}=
z^k+\sum_{j>0}\left(\wid{H}^{2k}_{0,j}z^{-j}+
\wid{H}^{2k}_{1,j}\theta z^{-j}\right) \\
\SH{2k+1}=
\theta z^k+\sum_{j>0}\left(\wid{H}^{2k+1}_{0,j}z^{-j}+
\wid{H}^{2k+1}_{1,j}\theta z^{-j}\right)
\end{array}\right.,$$
with $\SH{0}=1$, $\overline{\SH{k}}=k\bmod 2$, and let
$$W=\Span_\Lambda\{\SH{k},\ k\ge 0\}\subset V.$$
It is not difficult to realize that $M$ is isomorphic to the big cell 
of the $0\vert 0$ component of $SGr_\Lambda$ (i.e. the open subset 
where $\ker\pi_W=\coker\pi_W=0$),
an explicit isomorphism being given by the map
$$\{\SH{k}\}_{k\ge 0}\mapsto\Span_\Lambda\{\SH{k},\, k\ge 0\}.$$
The {\em super central system} is the dynamical system defined on $M$ by
requiring that
$$(\de_{t_j}+\SH{j})\cdot W\subset W$$
or, equivalently,
\begin{equation}\label{eq:bi}
\de_{t_j}\SH{k}=-\pi_-(\SH{j}\SH{k}),
\end{equation}
where $\pi_-:V\to V_-$ is the projection of $V=V_-\oplus V_+$ onto $V_-$
parallel to $V_+$.
\enddef
By comparing coefficients in the formulation~\rref{eq:bi} of the super central
system, we can explicitly write its evolution equations (see Table 2).
Notice that the SCS is to be thought of as a system of $\ZZ_2$-graded {\em
  ordinary} differential equations. 
\begin{center}
\begin{figure}[hbt]
\begin{tabular}{|| rl ||}
\hline\hline
 & \\
Table 2: & The SCS equations\\
 & \\
\hline\hline
 & \\
$\de_{t_{2k}}\wid{H}^{(2j)}=$ &
$\wid{H}^{(2j+2k)}-\wid{H}^{(2k)}\wid{H}^{(2j)}+
\dsl{\sum_{l=1}^j\left(\wid{H}^{2k}_{0,l}\wid{H}^{(2j-2l)}+
\wid{H}^{2k}_{1,l}\wid{H}^{(2j-2l+1)}\right)}$ \\ 
 & $\qquad+\dsl{\sum_{l=1}^k\left(\wid{H}^{2j}_{0,l}\wid{H}^{(2k-2l)}+
\wid{H}^{2j}_{1,l}\wid{H}^{(2k-2l+1)}\right)}$
\\ &\\ 
$\de_{t_{2k}}\wid{H}^{(2j+1)} =$ &
$\wid{H}^{(2j+2k+1)}-\wid{H}^{(2k)}\wid{H}^{(2j+1)}+
\dsl{\sum_{l=1}^j\wid{H}^{2k}_{0,l}\wid{H}^{(2j-2l+1)}}$ \\ 
& $+\dsl{\sum_{l=1}^k\left(\wid{H}^{2j+1}_{0,l}\wid{H}^{(2k-2l)}+
\wid{H}^{2j+1}_{1,l}\wid{H}^{(2k-2l+1)}\right)}$ \\ &\\ 
$\de_{t_{2k+1}}\wid{H}^{(2j)} =$ &
$\wid{H}^{(2j+2k+1)}-\wid{H}^{(2k+1)}\wid{H}^{(2j)} 
+\dsl{\sum_{l=1}^k\wid{H}^{2j}_{0,l}
\wid{H}^{(2k-2l+1)}}$ \\
& $ +\dsl{\sum_{l=1}^j\left(\wid{H}^{2k+1}_{0,l}\wid{H}^{(2j-2l)}+
\wid{H}^{2k+1}_{1,l}\wid{H}^{(2j-2l+1)}\right)}$\\ & \\ 
$\de_{t_{2k+1}}\wid{H}^{(2j+1)}=$&
$-\wid{H}^{(2k+1)}\wid{H}^{(2j+1)}+\dsl{\sum_{l=1}^j\wid{H}^{2k+1}_{0,l}
\wid{H}^{(2j-2l+1)}-
\sum_{l=1}^k\wid{H}^{2j+1}_{0,l}\wid{H}^{(2k-2l+1)}}$\\ &\\ & \\
\hline\hline
\end{tabular}
\end{figure}
\end{center}
\subsection{HSKP as a ``reduction'' of SCS}\label{sect:scs-hskp}
The Hamiltonian super KP hierarchy~\rref{eq:defHSKP} can be obtained 
from SCS by ``spatialisation''.  This procedure will be used 
in Section~\ref{linear} to produce solutions of HSKP starting from 
solutions of SCS. 
A spatialization of a \ger y of dynamical systems $X$ is
a process, (see, e.g., ~\cite{CS}), consisting 
in the projection of $X$ onto the space 
$\wid{\cal Q}_j$ of solutions of its $j$-th flow.
More informally, it boils down  to interpret
a distinguished flow parameter as a space coordinate, and allows to interpret
the dynamical system as a system of PDEs. 

In the ordinary KP case, 
spatialization with respect to the time $t_1=x$  simply
amounts to {\em identify} $t_1$ with $x$ (or better, substitute $t_1=x+t_1$ in
the solutions of CS). This procedure yields that
$h(x)=\Ha{1}_{\vert_{t_1=t_1+x}}$ is a solution to the KP equations.   

In the super case,  we want to consider the projection 
of SCS to the space $\wid{\cal Q}_2$
of solutions of its second flow i.e. the space of orbits of $\de_2$.
Essentially, we have to consider $k=1$ and interpret the first 
two families of equations of motion
reported in Table 2 as recursive definitions of the currents, as
differential polynomials (in the space variable $x=t_2$) of the generators
$\widehat{H}^{(1)}$ and $\widehat{H}^{(2)}$.
With respect to the bosonic case, there is a subtlety, 
connected with the relation of the first time $t_1$ of
SCS with the fermionic partner $\varphi$ of $x$. 
Observe that, by the definition of the super central system,
we have $\de_{t_j}\SH{k}=(-1)^{jk}\de_{t_k}\SH{j}$.
Notice in particular that $\de_1\SH{1}=0$.
Now, for $k>1$
\begin{equation}\label{eq:dop}
(-1)^k(\de_1+t_1\de_2)\SH{k}=\de_{t_k}(\SH{1}+t_1\SH{2}),
\end{equation}
suggesting that in order to get HSKP (and solutions thereof) we should put
$\hat{h}=\SH{1}+t_1\SH{2}$, $t_2=x$ and $t_1=\varphi$.
This is ``almost true'', but one must pay attention to the {\em order} in
which these identifications are performed. Indeed, 
plugging $k=1$ in the left hand side of~\rref{eq:dop}, we get
$$-(\de_1+t_1\de_2)\SH{1}=-t_1\de_1\SH{2}\ne
\de_1(\SH{1}+t_1\SH{2}),$$
which is unconsistent.

The right way to proceed is the following. Starting from a solution of SCS,
which depends on the times $(t_1,t_2,\ldots)=\mathbf t$,
one first replaces in the currents $\SH{j}(\mathbf t)$ 
the times $t_1$ with $t_1+\varphi$ and $t_2$ with $x$,
then one defines
\begin{equation}\label{eq:repla}
\hat{h}(x,\varphi;\mathbf t):=\SH{1} (t_1+\varphi,x,\ldots)+
\varphi  \SH{2}(t_1+\varphi,x,\ldots).
\end{equation}
Since $\de_1\SH{k}=\de_\varphi\SH{k}$ for any $k$, and taking also into 
account~\rref{eq:dop}, now we  have that the field 
$\hat{h}(x,\varphi;\mathbf t)$ is a solution of
$\de_{t_k}\hat{h}(x,\varphi;\mathbf t)=(-1)^k\delta\widehat{H}^{(k)}$, 
i.e. that it indeed satisfies the HSKP hierarchy.
Observe that $\SH{2}=\hat{h}^{(2)}=(\delta\hat{h})$; in fact
\begin{eqnarray*}
\hat{h}^{(2)} & = & \delta(\wid{H}^{(1)}+\varphi\wid{H}^{(2)})=
\de_\varphi\wid{H}^{(1)}+\varphi\de_x\wid{H}^{(1)}+
\de_\varphi(\varphi\wid{H}^{(2)}) \\
& = & \varphi\de_\varphi\wid{H}^{(2)}+(\wid{H}^{(2)}-
\varphi\de_\varphi\wid{H}^{(2)})=\wid{H}^{(2)}.
\end{eqnarray*}
Finally one has  that
$(\de_{t_j}+\SH{j})W\subset W$ implies
$(\delta+\hat{h})W_{B_{[x\varphi]}}\subset W_{B_{[x\varphi]}}$.
\subsection{The connection with the JSKP of Mulase and Rabin}\label{others}
In this section we will show that HSKP is equivalent to the Mulase--Rabin
Jacobian Super KP \ger y. Although
it seems conceivable from the supercommutativity
of the flows, such an identification is somewhat subtle. 
Our essential tool will be the introduction of {\em two} wave or 
Baker--Akhiezer functions associated with HSKP.\\
The zero curvature condition~\rref{eq:0cur} implies the existence of a first
wave function $\Phi$ satisfying
\begin{equation}\label{eq:defPhi}
\dpt{}{k}\Phi=\SH{k}\Phi.
\end{equation}
Now we perform the following ``trick'', whose meaning
will be discussed in Remark~\ref{rema21}. We define 
``enhanced'' currents $\GH{j}$ by the formula:
\begin{equation}\label{eq:Kdef}
\GH{j}=\SH{j}+(-1)^{j+1}\varphi \oint \SH{j} dz\,d\theta\>.
\end{equation}
In words, the difference between $\GH{j}$ and $\SH{j}$ is (up to a sign)
the $\theta$ component of the residue in $z$ of $\SH{j}$, multiplied by
$\varphi$; in the sequel we will denote it as
\[
\GH{j}-\SH{j}=(-1)^{j+1}\varphi\,C_j\>.
\]  
The zero curvature condition  on the currents $\SH{j}$
implies that the enhanced
currents satisfy the analogue condition
\begin{equation}
  \label{eq:0curK}
  \dpt{\GH{j}}{n}=(-1)^{j\,n}\dpt{\GH{n}}{j},
\end{equation}
and so guarantees the existence of an enhanced 
wave function $\Psi$ satisfying
\begin{equation}\label{eq:defPsi}
\dpt{}{n}\Psi=\GH{n}\Psi.
\end{equation}
The wave function $\Psi$ is readily seen to be related to the $\Phi$--wave
function by the formula:
\begin{equation}
  \label{eq:phipsi}
  \Psi=\Phi\cdot \exp\Big(-\varphi\int_{\mathbf{t}^{ev}_0}^{\mathbf{t}^{ev}} 
    \sum_n C_{2n} d s_{2n}\Big).
\end{equation}
We now consider the logarithmic derivative
\begin{equation}
  \label{eq:kappa}
  \ggh=\delta\Psi/\Psi.
\end{equation}
It is related with the \Fdb\ generator $\hat{h}$ by
\begin{equation}
  \label{eq:difhk}
  \ggh=\hat{h}-\int_{\mathbf{t}^{ev}_0}^{\mathbf{t}^{ev}}\sum_n 
C_{2n} ds_{2n}\>.
\end{equation}
Actually, since $C_2$ is readily seen to be $\psi_1$, we can say that
$\ggh$ is a $\CC^*[z,\theta]$--valued superfield of the form
\begin{equation}
  \label{eq:formak}
  \ggh=\theta a+\tilde{\nu}+\varphi h +(\theta\varphi)\, \psi,
\end{equation}
where now
\begin{equation}
  \label{eq:nubar}
  \tilde{\nu}=\nu_0+\sum_{j\ge 1} \frac{\nu_j}{z^j},\> \mbox{ with } 
        \nu_{0x}=-\psi_1. \end{equation}
We notice that, by a straightforward supersymmetric
extension of  standard properties of
the \Fdb\ procedure, since $\ggh$ differs from $\hat{h}$ by a zero order
term in $z$, one can write the \Fdb\ iterates $\gh{j}$ of $\ggh$ as a
linear (over $\Bring$) combination of the iterates $\sh{j}$ we have been using 
so far (and conversely).  To grasp this fact, one simply has to notice that
Lemma~\ref{lem21} holds irrespectively of the fact that $\nu_0$
vanishes. Since $\hat{f}^\prime=\delta\ggh=\hat{f}+\delta\nu_0$, we have 
(using induction) that
\[\begin{split}
(\del_x+\hat{f}^\prime)&\sum \alpha_i\sh{i}=\big(\del_x+\hat{f}
+\delta\nu_0\big)
(\sum\alpha_i\sh{i})\\ &=\sum \alpha_i\sh{i+2}+\sum \alpha_{ix}\sh{i}+\sum
(\delta\nu_0\alpha_i)\sh{i}.\end{split}
\] 
Summing up, we see that we can express the enhanced currents as linear
combinations of the \Fdb\ iterates of $\ggh$:
\[
\GH{j}=\sum_m \Gamma_m^j \gh{m}.
\]
Thanks to the obvious equality $\gh{j}\Psi=\delta^j\Psi$, we can write 
equation~\rref{eq:defPsi} as follows:
\begin{equation}\label{eq:jhskp}
\del_{t_j}\Psi=\GH{j}\Psi=\sum_l\Gamma^j_l\gh{l}\Psi=B_j\cdot\Psi\>,
\end{equation}
where $B_j$ is a super differential operator of order $j$ and of parity
$j\bmod 2$.

This equation is the bridge between HSKP and JSKP.
The latter is usually 
formulated within the theory of super pseudo-differential operators. 
A discussion of such a topic is outside the size 
of this paper (see, e.g.,  \cite{MaRad,SolvBirk,Mu} for details).  
We need the following lemma, whose proof is a straightforward computation:
\begin{lemma}\label{lemS} Let $S$ be a super pseudo-differential operator, with
  coefficients in $\Bring$ of the form
\begin{equation}\label{eq:ss}
S=1+\sum_{j>0}\big(u_j+\varphi\xi_j\big)\del_x^{-j}+\big(\eta_j+\varphi
w_j\big)\delta^{-(2j-1)}.
\end{equation}
and let
\[
e(z,\theta;x,\varphi,{\bf
    t})=\exp\big(\varphi\theta+zx+\sum_{j>0}(t_{2j}z^j+t_{2j-1} \theta
  z^{j-1})\big)
\]
be the vacuum wave function (in the terminology of~\cite{IbMaMe96}) for JSKP.
Then, if $\Psi$ is a Baker--Akhiezer function for JSKP obtained by dressing
with $S$ the vacuum wave function $e$, 
\begin{equation}
  \label{eq:dress2}
  \Psi=S\cdot e,\qquad 
\end{equation}
its logarithmic derivative $\delta\Psi/\Psi$ is a superfield  of the
form~\rref{eq:formak}, satisfying the constraint $\nu_{0x}+\psi_1=0$. 
\end{lemma}
The JSKP equations can now be obtained by means of standard procedures in the 
theory of integrable systems. Indeed,
taking the $t_{2j}$ and $t_{2j-1}$ derivatives of the dressing
relation~\rref{eq:dress2}, and taking also equation~\rref{eq:jhskp} into
account,  we have
\begin{equation}\begin{split}
\de_{t_{2j}}\Psi & =  \de_{t_{2j}}(S\cdot e)=(\de_{t_{2j}}S)\cdot e
+S\cdot z^{j}e 
=  (\de_{t_{2j}}S)\cdot e+S\delta^{2j}\cdot e \\
& =  \left((\de_{t_{2j}}S)S^{-1}+S\delta^{2j}S^{-1}\right)S\cdot e 
 =  \left((\de_{t_{2j}}S)S^{-1}+S\delta^{2j}S^{-1}\right)\cdot\Psi=
B_{{2j}}\cdot\Psi\end{split}
\end{equation}
and
\begin{equation}\begin{split}
\de_{t_{2j-1}}\Psi & = \de_{t_{2j-1}}(S\cdot e)=(\de_{t_{2j-1}}S)\cdot e+
S\cdot \theta z^{j-1}e 
 =  (\de_{t_{2j-1}}S)\cdot e+S(\delta^{2j-1}-\varphi \delta^{2j})\cdot e \\
& =  \left((\de_{t_{2j-1}}S)S^{-1}+S(\delta^{2j-1}-\varphi \delta^{2j})S^{-1}
\right)\cdot\Psi 
=  B_{{2j-1}}\cdot\Psi.
\end{split}
\end{equation}
Since $(\de_jS)S^{-1}=((\de_jS)S^{-1})_-$ is a purely pseudo-differential
operator (i.e. it has no differential part) we get
$$\de_{t_{2j}}S=-(S\delta^{2j}S^{-1})_-S=-(S\de^j_xS^{-1})_-S$$
and
$$\de_{t_{2j-1}}S=-(S(\delta^{2j-1}-\varphi \delta^{2j})S^{-1})_-S=
-(S\de_\varphi\de^{j-1}_xS^{-1})_-S,$$
which are the equations that Mulase and Rabin defined for JSKP.
\brem \label{rema21}
As we have anticipated, the introduction of the enhanced currents
  $\GH{j}$ is not a mere trick. To better understand their origin it
is useful to reconsider our choices, namely the
decomposition~\rref{eq:deco1} of the space $V_\Bring$ of formal 
Laurent series with coefficients in $\Bring$ as the direct sum of the 
subspace generated by the \Fdb\ monomials $W_{B_{[x\varphi]}}$ and
the space   $V_{B_{[x\varphi]}}^-$  of formal power series without 
``constant term'',
i.e. starting from $z^{-1}$ and $\theta z^{-1}$. If $\pi_+: V_\Bring\to
W_{B_{[x\varphi]}}$ is the projection associated with such a decomposition, 
then
the currents $\SH{k}$ are given by the formulas
\[
\SH{2j+p}=\pi_+(z^j\theta^p),\> j\in \NN,\> p\in \{0,1\}.
\]
Actually, associated with our geometrical datum, there is another
natural choice. Indeed, one simply notice the fact that
it is possible to extend
the \Fdb\ recursion relations~\rref{eq:sfdbrecrel} to negative values of the
index $j$, and get a full \Fdb\ basis $\big\{\gh{j}\big\}_{j\in\ZZ}$ in  
$V_{B_{[x\varphi]}}$. The asymptotics of the \Fdb\ basis is readily seen to be
\begin{equation}\label{eq:fdbasy}
\gh{2j}\sim z^j,\quad
\gh{2j-1}\sim \theta z^{j-1}+\varphi z^{j}.
\end{equation}
Hence we have another natural decomposition
\begin{equation}\label{eq:deco2}
V_{B_{[x\varphi]}}=W_{B_{[x\varphi]}}\oplus W_{B_{[x\varphi]}}^-,
\end{equation}
where now $W_{B_{[x\varphi]}}^-$ is the space generated by the \Fdb\ iterates
with  {\em negative} index. If we call
\[
{\pi}_+^\prime: V_\Bring\to W_{B_{[x\varphi]}}
\]
the projection associated with this new decomposition, then we have that the
enhanced currents are given by
\[
\GH{2j+p}=\pi_+^\prime(z^j\theta^p),\> j\in \NN,\> p\in \{0,1\}.
\]    
Actually, it is not surprising that the 
connection with the usual formulation of the
theory by means of super pseudo-differential operators can be better described 
using the decomposition associated with the full \Fdb\ basis, since
$\delta^j\Psi=\gh{j}\Psi$ and the projection $\pi_+^\prime$ is adapted to 
the projection which kills the non-differential part of a super 
pseudodifferential operator.
\erem
\brem 
It is actually possible to write an evolution
  equation of the form ~\rref{eq:jhskp} for the first wave function $\Phi$ as
  well. Indeed, since $\hat{h}\Phi=\delta\Phi$ and $\SH{j}$ is a linear
combination of the \Fdb\ iterates $\sh{j}$ of $\hat{h}$, we can read the
equation $\del_{t_j}\Phi=\SH{j}\Phi$
as $\del_{t_j}{\Phi}=B_j^\prime\Phi$
where $B_j^\prime$ is still  a super differential operator 
of order $j$ and of parity
$j\bmod 2$. However, we can no more express $\Phi$ as the result of the 
action of a {\em  generic} dressing operator $S$ on 
the vacuum wave function $e$ of Lemma~\ref{lemS}.
Moreover, the HSKP equations are not compatible with $\Phi$ having such
an expression: even if $\psi_1=0$ at $\mathbf{t}_0$, this is
no more true for the evolved field so $0=\nu_{0x}\ne-\psi_1$.\erem
\brem
As we have seen, the connection between HSKP and JSKP is (albeit in a
  tricky way) a change of coordinates. 
  The relation among the degrees of freedom
  $(u_i,w_i,\xi_i,\eta_i)$ of $S$ and the degrees of freedom
  $(a_i,h_i,\nu_i,\psi_i)$ (collected in $(a(z),h(z),\nu(z)\psi(z))$ as usual)
of $\ggh$ is indeed the following:
\begin{equation}
  \label{eq:coordchange}
  \begin{array}{l}
\nu(z)\Big(1+\sum_{i>1}\dsl{
\frac{u_i}{z^i}}\Big)=-\eta_1+\sum_{i>1}\dsl{\frac{\xi_i-\eta_{i+1}}{z^i}};\\
a(z)\Big(1+\sum_{i>1}\dsl{\frac{u_i-\eta_i\nu(z)}{z^i}}\Big)=1+\sum_{i>1}\dsl{
\frac{u_i+w_i}{z^i}};\\
h(z)\Big(1+\sum_{i>1}\dsl{\frac{u_i}{z^i}}\Big)=z+u_1+\eta_1\nu(z)
+\sum_{i>1}\dsl{\frac{u_{i,x}+u_{i+1}}{z^i}}\\
\qquad\qquad+\nu(z)\Big(\sum_{i>1}\dsl{\frac{\xi_i-\eta_{i+1}}{z^i}}\Big);\\
\psi(z)\Big( 1+\sum_{i>1}\dsl{ \frac{u_i}{z^i}}\Big)=a(z)\Big(\eta_1+\sum_{i>1}
\dsl{\frac{\eta_{i+1}-\xi_i}{z^i}}\Big)+\nu(z)\sum_{i>1}\dsl{\frac{w_i}{z^i}}\\
\qquad\qquad -h(z)\sum_{i>1} \dsl{\frac{\eta_i}{z^i}}
+\sum_{i>1} \dsl{\frac{\eta_{ix}+\xi_i}{z^i}}.
\end{array}
\end{equation}
These equations give   $(a_i,h_i,\nu_i,\psi_i)$ as differential polynomials in 
the $(u_i,w_i,\xi_i,\eta_i)$'s, and can be inverted modulo quadratures (as
usual in the theory of KP--like  equations). As we shall show in the next
sections, there is some merit in considering such non standard coordinates, 
whose choice is suggested by the supersymmetric extension of
Gel'fand--Zakharevich set-up for the KP theory.
\erem
\subsection{A super KdV equation as a reduction of HSKP}\label{sect:1-1}
We will discuss now a supersymmetric generalization of the
KdV equation obtained as reduction of the Hamiltonian super KP hierarchy. 
This example will be important in giving us one more clue to 
the second part of the paper. 
It can be shown that constraints of the form
\[
\SH{2k}=z^k 
\]
are compatible with HSKP. We consider $k=2$ obtaining
\begin{equation}\label{eq:KDV-constraints}
\begin{array}{l}
\left\{\begin{array}{l}
h_2=-\frac{1}{2}h_1'+\psi_1\nu_1 \\
h_k=-\frac{1}{2}h_{k-1}'-\frac{1}{2}\sum_{j=1}^{k-2}h_{k-j-1}h_j+
\psi_1\nu_{k-1}
\qquad\mbox{for $k>2$,}
\end{array}\right.\\
\left\{\begin{array}{l}
\psi_2=-\frac{1}{2}\psi_1'+a_1\psi_1 \\
\psi_k=-\frac{1}{2}\psi_{k-1}'-\sum_{j=1}^{k-2}h_{k-j-1}\psi_j+
a_{k-1}\psi_1
\qquad\mbox{for $k>2$,}
\end{array}\right. \\
\left\{\begin{array}{l}
\nu_2'=-\frac{1}{2}\nu_1''+a_1'\nu_1 \\
\nu_k'=-\frac{1}{2}\nu_{k-1}''-\sum_{j=1}^{k-2}h_{k-j-1}\nu_j'+
a_1'\nu_{k-1}
\qquad\mbox{for $k>2$,}
\end{array}\right. \\
\left\{\begin{array}{l}
a_2'=-\frac{1}{2}a_1''+a_1a_1' \\
a_k'=-\frac{1}{2}a_{k-1}''-\sum_{j=1}^{k-2}h_{k-j-1}a_j'+a_{k-1}a_1'
\qquad\mbox{for $k>2$}\>.
\end{array}\right.
\end{array}
\end{equation}
These equations allow us to compute recursively the coefficients $h_j$,
$\psi_j$, $\nu_j$ and $a_j$ for $j>1$ in terms of $h_1$, $\psi_1$, $\nu_1$
and $a_1$ by means of quadratures.
In order to explicitly write the equations for the independent 
degrees of freedom we have
only to calculate the coefficients of order $1$ of the super current
densities. Since the relations~\rref{eq:KDV-constraints}
algebraically determine only the derivatives of the
fields $a_i,\nu_i,\i\ge 2$, 
in general the resulting equations will be
integro--differential ones. Fortunately enough, for the sixth time of 
the \ger y the non local terms cancel each other, and the result is
\begin{equation}\label{eq:skdv}
\left\{\begin{array}{l}
\de_6\nu_1=\frac{1}{4}\nu_1'''-\frac{3}{2}a_1'\nu_1'-3h_1\nu_1' \\
\de_6a_1=\frac{1}{4}a_1'''-\frac{3}{2}{a_1'}^2-3h_1a_1'+6\psi_1\nu_1' \\
\de_6h_1=\frac{1}{4}h_1'''-3h_1h_1'-
\frac{3}{2}\psi_1\nu_1''-\frac{3}{2}\psi_1'\nu_1' \\
\de_6\psi_1=\frac{1}{4}\psi_1'''-\frac{3}{2}a_1''\psi_1-
\frac{3}{2}a_1'\psi_1'-
3h_1\psi_1'-3h_1'\psi_1
\end{array}\right. \>.
\end{equation}
We thus see that the evolution equations for 
the time $t_6$ are a supersymmetric
extension of the KdV equation, which can be retrieved by setting
$a_1=\nu_1=\psi_1=0,h_1=\dsl{\frac{u}{2}}$.
We notice also the following fact. 
Substituting $\nu=\psi=0,h=\dsl{\frac{u}{2}}$
  in the above equations~\rref{eq:skdv} we obtain the ordinary system of PDE's 
  in two variables $u$ and $a$
\begin{equation}\label{eq:mkdv}
\left\{\begin{array}{l}
\del_t{a}=\frac{1}{4}a_{xxx}-{\frac{3}{2}}{a_{x}}^2-\frac32 u a_{x}\\
\del_t{u}=\frac{1}{4}u_{xxx}-\frac{3}{2}u u_{x}
\end{array}\right. \>.
\end{equation}
One can easily notice that the submanifold defined by $u=-a_{x}+a^2+\lambda_0$
is an invariant submanifold of these equations, where the first one coincides
with the {\em modified} KdV equation. 
So we see that this reduction of HSKP ``contains'' both KdV and 
mKdV. This observation will be formalized and explained in the next sections.
\section{HSKP and Darboux transformations}\label{Darboux}

In general, a Darboux transformation is a way to connect two
systems of differential equations enabling to produce a solution of the
second once a solution of the first has been supplied.
This technique has proved to be very effective
both in the construction of large classes of explicit solutions of soliton
equations and in the understanding of the nature of infinite dimensional
integrable systems~(see, e.g., \cite{MS,K1}).
An example of such transformation is provided by the Miura map in the KdV
theory and the modified KdV hierarchy (mKdV)
(see, e.g., \cite{D,MS} and the references quoted therein).
Here we are mostly interested in the concept of Darboux intertwiners and
Darboux coverings introduced in \cite{MPZ,MZ}, where the geometrical features
of the method where analyzed as follows.
 
Consider three vector fields $X$, $Y$ and $Z$ on three manifolds $M$, $N$ and
$P$, respectively.
\begindef\label{covering}\cite{MPZ}
The vector field $Y$ {\em intertwines} $X$ and $Z$ if there exists
a pair of maps $(\mu:N\to M,\sigma:N\to P)$ such that $X=\mu_*Y$ and
$Z=\sigma_*Y$.
Moreover, if $X=Z$ , $N$ is a fiber bundle  on $ M=P$,  and $\mu:N\to M$ is the
bundle projection, then $Y$ is said to be
a {\em Darboux covering} of $X$ , and the map $\sigma$ the associated {\em
  Miura} map.
\begin{figure}[htb]
  \caption{The Darboux Maps}
\bigskip\bigskip
\centerline{\epsfxsize=6in\epsfbox{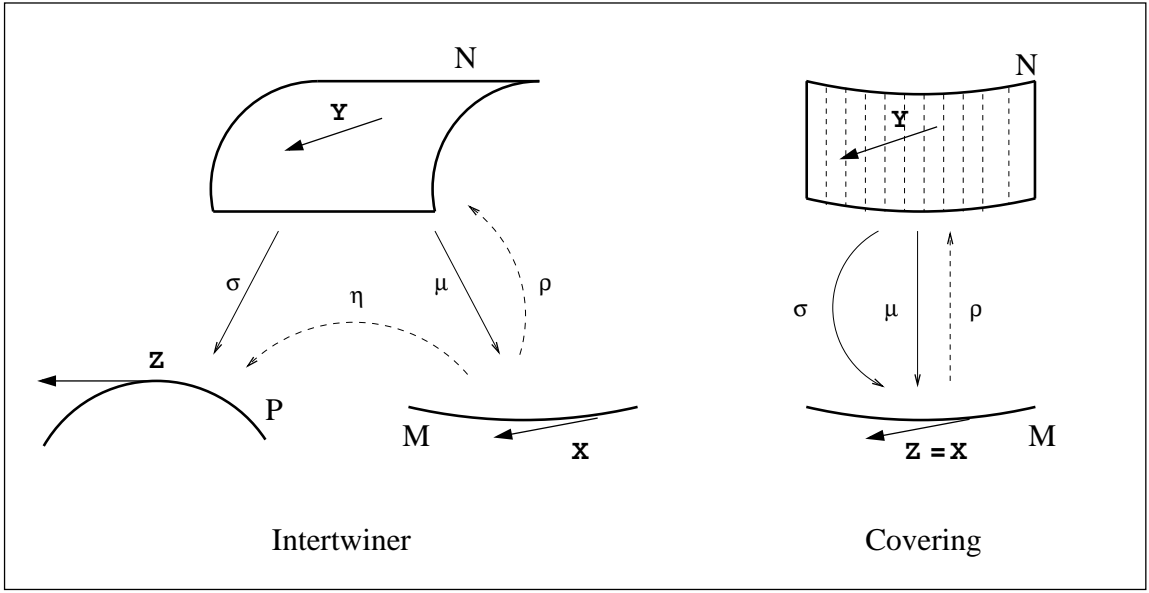}}
 \end{figure}
Finally,  still when $X=Z$ and $\mu:N\to M$ is a fibre bundle, for each
section $\rho:M\to N$ of $\mu$ which is  invariant under $Y$, 
the composition $\eta=\sigma\circ \rho$ which 
sends $X$ in $X$, and hence produces 
an integral curve $\tilde{x}(t)$ of $X$ from
the integral curve $x(t)$ of $X$, is called a {\em Darboux transformation}.
\enddef
The concept of Darboux covering 
is useful for constructing both solutions and invariant
submanifolds of the vector field $X$: if $U$ is a chart on $M$ with coordinate
$x$ and $V\subset\mu^{-1}(U)$ a chart on $N$ adapted to the projection $\mu$
and with fibered coordinates $(x,a)$, then the local expression of the above
vector fields is
$$\begin{array}{l}
\dot{x}=X(x) \\
\dot{a}=Y(x,a),
\end{array}$$
where the first equation is that of $X$ on $U$.
Then, any integral curve $x(t)$ of $X$ can be lifted to an integral curve
$(x(t),a(t))$ of $Y$ by solving the second equation, which is controlled by
$x(t)$.
Therefore, we get a new integral curve of $X$ by setting
$$\tilde{x}(t)=\sigma(x(t),a(t)).$$
The last equation can also be interpreted as a ``symmetry (or Darboux) 
transformation''
of the dynamical system described by $X$, depending on a solution of the
auxiliary system for $a$, controlled by $X$ itself, which associates
$\tilde{x}(t)$ with $x(t)$. 

The application of the formalism we have just described to KP naturally leads
to the DKP hierarchy.
\begindef\label{defDKP}
Let $M$ be the affine space of (formal) monic Laurent series in $z^{-1}$ with
coefficients in $C^\infty(S^1)$ and of the form
$$h(z,x)=z+\sum_{j>0}h_j(x)z^{-j}$$
and let $N$ be the affine space of couples $(h,a)$ where $h$ is as above and
$a$ is a monic Laurent series of the form
$$a(z,x)=z+\sum_{j\ge 0}a_j(x)z^{-j}.$$
Define two maps $\mu,\sigma:N\to M$ by
$$\mu(h,a)=h$$
and
$$\tilde{h}:=\sigma(h,a)=h+\frac{\de_x a}{a}.$$
Finally, let $H^{(k)}$ and $\tilde{H}^{(k)}$ be the current densities
associated with  $h$ and $\tilde{h}$, respectively.
The {\em DKP hierarchy} is the hierarchy of evolution equations on $N$
defined by
$$\left\{\begin{array}{l}
\frac{\de}{\de t_k} h=\de_xH^{(k)}\\
\frac{\de}{\de t_k}a=a(\tilde{H}^{(k)}-H^{(k)})
\end{array}\right. .$$
\enddef
DKP is a Darboux covering, in the sense of Definition \ref{covering}, of the
KP hierarchy
$$\de_{t_k}h=\de_xH^{(k)}.$$
Indeed, it is clear that $\mu_*$ maps the vector fields $\de_{t_j}$ of DKP to
those of KP.
As for $\sigma_*$, we have
\begin{equation}
\begin{split}
\de_{t_k}\left(\frac{\de_x a}{a}\right) & = 
\frac{a\de_x\de_{t_k}a-(\de_xa)(\de_{t_k}a)}{a^2} \\ 
= & \frac{a\de_x(a\tilde{H}^{(k)}-aH^{(k)})-
a(\de_xa)(\tilde{H}^{(k)}-H^{(k)})}{a^2} \\
= & \de_x\tilde{H}^{(k)}-\de_xH^{(k)}\end{split}
\end{equation}
and finally
\begin{equation}
\de_{t_k}\tilde{h}=\de_{t_k}h+\de_{t_k}\left(\frac{\de_x a}{a}\right)=
\de_x\tilde{H}^{(k)}.
\end{equation}
In the papers~\cite{MZ,MPZ,Rat,FMP}, the following results 
were obtained:
\begin{enumerate}
\item\label{it1} The modified KP hierarchy of ~\cite{Kup} 
is the restriction of DKP on the invariant submanifold
$S_0\subset N$ defined by the simple equation $a=h+a_0$;
\item\label{it2} The DKP equations admit a remarkable family 
of invariant submanifolds,
  $S_l$, of which $S_0$ is the simplest; the images through $\mu$ of the
  intersections of {\em two} (or more) submanifolds, $S_{l_1}\cap\cdots\cap
  S_{l_k}$ is an invariant submanifold of KP, which coincide with the
  rational KP reductions of Dickey and Krichever (see, e.g.,
  \cite{Dik94,Kr95,Ar95,BoLiXi95});
\item\label{it3} The central system CS  can be explicitly linearized and 
classes of solutions can be explicitly found  by means  of a Darboux 
intertwiner linking it  with the Sato system, that is, the coordinate 
expression of the linear flows of KP on the Sato Grassmannian.    
\end{enumerate}
In the rest of this Section we will first show that HSKP (and hence JSKP) can
be seen as a supersymmetric extension of DKP; then  we will define a Darboux
covering for HSKP, and briefly discuss the analogue of the invariant
submanifolds mentioned in points~\ref{it1} and \ref{it2} of the above list.
The generalization of point~\ref{it3} will be the subject of
Section~\ref{linear}.
\subsection{HSKP and Darboux transformations}\label{JSKP-Darboux}

Our first goal is to give a  connection between the
Jacobian super KP hierarchy and DKP.
First of all, we observe that the role $a$ has in DKP does not depend
on the order of its pole, since it appears in a homogeneous way in all
the equations.
Hence, we see that (the reduction modulo nilpotents elements in $\La$ of) 
the bosonic degrees of freedom of HSKP are exactly the
degrees of freedom of $DKP$: the Laurent series $a$ appearing
in the definition~\rref{eq:hath} of the super \Fdb\ generator 
can be identified with  $z^{-1}$ times the Laurent series $a$ appearing in the
definition~\ref{defDKP} of DKP.
It is thus tempting to conjecture a relation between the two hierarchies.
In fact, we can prove the following
\begin{proposition}
Let $\hat{h}$ and $\SH{k}$ be, respectively, the super \Fdb\ generator
and the currents of HSKP defined as in Section \ref{FdBSKP}.
\begin{itemize}
\item[i.] The constraint $\nu=\psi=0$ is compatible with the even flows of the
HSKP hierarchy.
\item[ii.] The reduction HSKP$_{bos}$ of the even flows of HSKP given by
setting $\nu=\psi=0$ is isomorphic to DKP, i.e. if $\hat{h}$ is a solution
of HSKP$_{bos}$, then $(h,za)$ is a solution of $DKP$ and vice versa.
\end{itemize}
\end{proposition}
\proof
\begin{itemize}
\item[i.] Looking at the recurrence relations we introduced in Section
\ref{FdBSKP}, namely equations~\rref{eq:formhh} and Table 1,
we see that, under the constraint $\nu=\psi=0$, one has
$$\left\{\begin{array}{l}
\hat{h}^{(2k-1)}=\theta a^{(k)}+\varphi h^{(k)} \\
\hat{h}^{(2k)}=h^{(k)}-(\theta\varphi)b^{(k)}\>,
\end{array}\right.$$
where the coefficients are given by the following recursion relations:
\begin{equation}\label{eq:redrecrel}
\begin{split} 
&\left\{\begin{array}{l}
h^{(k+1)}=(\de_x+h)h^{(k)} \\
h^{(0)}=1
\end{array}\right., \qquad
\left\{\begin{array}{l}
a^{(k+1)}=(\de_x+h)a^{(k)} \\
a^{(1)}=a
\end{array}\right.,\\
&\left\{\begin{array}{l}
b^{(k+1)}=(\de_x+h)b^{(k)}+(\de_xa)h^{(k)}
\\
b^{(0)}=0\>.
\end{array}\right.\end{split}
\end{equation}
In particular 
$\sh{0}=1,\> \sh{1}=\theta a+\varphi h,\> \sh{2}=h-\theta\varphi a_x.$
This implies that
$$\SH{2k}=H^{(k)}-(\theta\varphi)K^{(k)},$$
where $H^{(k)}$ is the $k$-th current density of KP and $K^{(k)}$ is some
power series in $z^{-1}$ and $x$ of the form
$$K^{(k)}(z,x)=\sum_{j>0}K^k_j(x)z^{-j}.$$
The evolution equations for the even flows of HSKP are then
$$\left\{\begin{array}{l}
\de_{t_{2k}}\nu=0\\ \de_{t_{2k}}\psi=0 \\
\de_{t_{2k}}a=K^{(k)} \\
\de_{t_{2k}}h=\de_xH^{(k)}\>, \\
\end{array}\right.$$
showing that the constraint $\nu=\psi=0$ is compatible with them.
\item[ii.] In the proof of i. we have established also that $h$ evolves
according to KP.
We need only to understand better the evolution of $a$.
We have to show that $K^{(k)}/a+H^{(k)}$ is the $k$-th current
density of KP associated with 
$$\widetilde{h}=h+{\frac{\de_xa}{a}}.$$
To achieve this we consider the function
\[ 
\wid{A}=\left(-\frac{1}{a}+\theta\varphi\right),
\] 
and perform the  ``gauge transformation''
$$\hat{h}^{(k)}\mapsto\hat{l}^{(k)}
:=\wid{A}\hat{h}^{(k)},$$ that is, we consider the new vector space
$\widetilde{W}=\wid{A}\cdot W$ generated by the $\hat{l}^{(k)}$'s.
We remark that the first transformed basis elements are:
\[\begin{split}
&\hat{l}^{(0)}=-\frac{1}{a}+(\theta\varphi)\cdot 1, \quad
\hat{l}^{(1)}=-\theta-\varphi\frac{h}{a} \\
&\hat{l}^{(2)}=-\frac{h}{a}+(\theta\varphi)\cdot
\left(h+\frac{\de_xa}{a}\right)=
-\frac{h}{a}+(\theta\varphi)\cdot \widetilde{h}.\end{split}
\]
Observe that since $W$ is generated by the action of the operator
$\del_x+\sh{2}$ on the pair $(\sh{0}=1,\sh{1}=\theta a+\varphi h)$,
$\widetilde{W}$ will be generated by the action of
$\wid{A}(\del_x+\sh{2})\wid{A}^{-1}$ on the pair $(\hat{l}^{(0)},
\hat{l}^{(1)})$.
We notice that
$$\wid{A}(\del_x+\sh{2})\wid{A}^{-1}=
\de_x+h+\frac{\de_xa}{a}=\de_x+\widetilde{h},$$
which shows that
\begin{equation}\label{eq:thephicomp}
\left\{\begin{split}
&\hat{l}^{(2k)}=f^{(k)}+\theta\varphi \widetilde{h}^{(k)}\\
&\hat{l}^{(2k+1)}=-\theta \widetilde{h}^{(k)}+ \varphi g^{(k)},
\end{split}\right.
\end{equation}

Now we consider the transformed current $\wid{L}^{(2k)}=\wid{A}\SH{2k}$. 
Its $\theta\varphi$ component is clearly given by the sum
\[
\wid{L}^{(2k)}_{\theta\varphi}=H^{(k)}+K^{(k)}/a\>.
\]
Since $\wid{L}^{(2k)}$ is a finite linear combination of the basis
elements $\hat{l}^{(j)}$ it follows that $\wid{L}^{(2k)}_{\theta\varphi}$ 
is the unique combination of the $\tilde{h}^{(k)}$ with the asymptotics
\[
 L^{(2k)}_{\theta\varphi}=H^{(k)}+K^{(k)}/a=z^k+O(1/z).
\]
Hence it must be equal to $\widetilde{H}^{(k)}$, so we get the desired 
result
\[
K^{(k)}=a(H^{(k)}-\tilde{H}^{(k)}).
\]
\end{itemize}
\endpf
\subsection{A Darboux covering for HSKP and the super analogue of its 
rational reductions}\label{sect:Dar2}
In this section we return to the full supersymmetric picture, define 
the  Darboux transformations and a D-HSKP hierarchy for the Hamiltonian 
super KP theory,
and show how to obtain the super analogue of Dickey's and Krichever's 
rational reductions of the KP \ger y.\\
We observe that given a Laurent series $\hat{h}$ of the usual form of
equation~\rref{eq:hath},
and a  monic even power series
$$\hat{p}=p+\theta\zeta+\varphi\xi+(\theta\varphi)q\>,$$
with $\bar{p}=\bar{q}=0$, $\bar{\zeta}=\bar{\xi}=1$ and
$$\left\{\begin{array}{l}
p=1+\sum_{j>0}p_jz^{-j} \\
q=\sum_{j>0}q_jz^{-j} \\
\zeta=\sum_{j>0}\zeta_jz^{-j} \\
\xi=\sum_{j>0}\xi_jz^{-j}\>,
\end{array}\right.$$
the transformed series $$\hat{k}=\hat{h}+\frac{\delta\hat{p}}{\hat{p}}\>.$$
is still of type~\rref{eq:hath}.
\begindef{\bf (D-HSKP)}\label{djskp}
Let $\wid{N}$ be the affine space of couples of monic formal Laurent series
$(\hat{h},\hat{p})$, let
$$\hat{k}=\hat{h}+\frac{\delta\hat{p}}{\hat{p}}$$
and let $\wid{K}^{(k)}$ be the $k$-th super current density associated to
$\hat{k}$.
The {\em Darboux--Hamiltonian super KP hierarchy} is the set of compatible
evolution equations
$$\left\{\begin{array}{l}
\de_{t_k}\hat{h}=(-1)^k\delta\SH{k} \\
\de_{t_k}\hat{p}=\hat{p}(\wid{K}^{(k)}-\SH{k})
\end{array}\right..$$
\enddef
If we let $\wid{M}$ be the affine space of the monic formal Laurent series
$\hat{h}$ and
we define two maps $\hat{\mu},\hat{\sigma}:\wid{N}\to\wid{M}$ by
$$\hat{\mu}(\hat{h},\hat{p})=\hat{h}$$
and
$$\hat{\sigma}(\hat{h},\hat{p})=\hat{h}+\frac{\delta\hat{p}}{\hat{p}},$$
then
\[
\begin{split}
\de_{t_k}\left(\frac{\delta\hat{p}}{\hat{p}}\right) & = 
(-1)^k\frac{\hat{p}\delta\de_{t_k}\hat{p}-(\delta\hat{p})(\de_{t_k}
\hat{p})}{\hat{p}^2} \\ 
= & (-1)^k\frac{\hat{p}\delta(\hat{p}\wid{K}^{(k)}-\hat{p}\SH{k})-
\hat{p}(\delta\hat{p})(\wid{K}^{(k)}-\SH{k})}{\hat{p}^2} \\
= & (-1)^k\delta(\wid{K}^{(k)}-\SH{k})\end{split}
\]
so
$$\de_{t_k}\hat{k}=\de_{t_k}\hat{h}+\de_{t_k}\left(\frac{\delta\hat{p} }
{\hat{p}}\right)= (-1)^k\delta\wid{K}^{(k)},$$
i.e. D-HSKP is a Darboux covering of HSKP.
In the next section we will use this formalism for a geometrical 
characterization of the analogue of the rational reductions of the KP \ger y.
\subsection{Super mKP and Rational Hierarchies}
\begin{proposition}
The submanifold $\wid{\cal S}_l$ of $\wid{N}$ (see Definition \ref{djskp})
characterized by
$$z^{l/2}\hat{p}\in W$$
for $l$ even, or by
$$\theta z^{(l-1)/2}\hat{p}\in W$$
for $l$ odd, is invariant under the flows of the D-HSKP hierarchy,
where we recall that  $W=\Span_\Bring\{\hat{h}^{(j)}\vert\ j\ge 0\}.$
Consequently, the submanifold
$$\wid{\cal T}_l:=\hat{\mu}(\wid{\cal S}_l)$$
of $\wid{M}$ is invariant under HSKP.
\end{proposition}
\proof
We give the proof only for $l=2n$ even, the other proof is the same up to
some obvious change of signs.
The condition $(\hat{h},\hat{p})\in\wid{\cal S}_l$ implies 
the existence of some
coefficients $\alpha_j(x,\varphi)$, $j=0,\cdots,l$ such that
\begin{equation}\label{eq:shatconstr}
z^n\hat{p}=\sum_{j=0}^l\alpha_j\SH{j},
\end{equation}
so we have to show that this expression is invariant under the flows of
D-HSKP, i.e.
\begin{equation}\label{ratrid}
\de_{t_k}\big(z^l\hat{p}-\sum_{j=0}^l\alpha_j\SH{j}\big)=0
\end{equation}  
on $\hat{\CS}_{2n}$.
Let $W^{\hat{k}}:=\Span_{\Bring}\{\hat{k}^{(j)}\vert\ j\ge 0\}$.
By definition we have
$$\hat{p}(\delta+\hat{k})=(\delta+\hat{h})\hat{p},$$
and hence $\hat{p}(\delta+\hat{k})^j=(\delta+\hat{h})^j\hat{p}.$
This implies that
$z^l\hat{p}\,W^{\hat{k}}\subset W,$
and, therefore, using the D-JSKP equations, 
$(\de_{t_k}+\SH{k})z^l\hat{p}=z^l\hat{p}\wid{K}^{(k)}\in W,$
i.e.
$$z^l\de_{t_k}\hat{p}+\sum_{j=0}^l(-1)^{jk}\alpha_j\SH{k}\SH{j}\in W.$$
Using now the property 
$\de_{t_k}\SH{j}+\SH{k}\SH{j}\in W$, characteristic of
HSKP, and comparing the coefficients of $z^j$ and $\theta z^j$ in
equation~\rref{eq:shatconstr} for
$j=0,\cdots,l$, we get
$$z^l\de_{t_k}\hat{p}-\sum_{j=0}^l(-1)^{jk}\alpha_j\de_{t_k}\SH{j}=
\sum_{j=0}^l(\de_{t_k}\alpha_j)\SH{j},$$
i.e. \rref{ratrid} holds.
\endpf
As a first application of this result we show how, in such a formalism, we
obtain a supersymmetric extension of the modified KP \ger y. We consider the
submanifold $\wid{\CS}_2$ defined by $z\hat{p}\in W$.
It can be seen that these equations entail the following constraints:
\begin{equation}
\zeta_1=0;\quad
\int_{S^1}\xi_j dx=0,\> j\ge 2;\quad
q_1=0;\mbox{ and }\int_{S^1}q_i dx=0\> j\ge 2.
\end{equation}
The bosonic sector of the resulting theory covers the
invariant submanifold $S_0$
of the DKP equations of~\cite{MPZ} defined by $a=h+a_0$. 
There it was proven that 
$DKP\vert_{S_0}$ is another form  of the modified KP theory of
Kupershmidt \cite{Kup}. Hence, through this result, 
we obtain that the restriction D--SKP${}_{\vert_{\CS_2}}$ provides 
a direct supersymmetric extension of mKP. 

Finally, following~\cite{Rat}, 
one can define and study the ``rational--type 
reductions'' of the HSKP \ger y as the restriction of the D-HSKP \ger y 
to the intersection of suitable of invariant submanifolds. In the next example
we will briefly describe the simplest case.
\bexample
Let us consider the {\em triple} intersection $\wid{\CS}_{124}$,
and its image $\wid{\CT}_{124}$ under $\hat\mu$,
obtained by requiring that the triple $(\theta\hat{p},z\hat{p},z^2\hat{p})$
lie in $W$. Since $\SH{1}=\theta a+\nu$ we see that, recalling the form of
$\hat{p}=p+\theta\zeta+\varphi\xi+(\theta\varphi)q$, the equation
$\theta\hat{p}\in W$ implies
\[
\nu=0,\quad \xi=0,\quad a=p.
\]
Intersecting with $\wid{\CS}_2$ we get
\[
h=zp-p_1,\quad \psi=-z\zeta+\zeta_1 p,\quad q=\frac{q_1p-p_x}{z}.
\]
Finally, requiring $z^2\hat{p}\in W$ one sees that it is possible to express
all the fields $p_i,q_i,\zeta_i,$ (and hence all the currents $\SH{j}$) in
terms of the two even fields
$r=p_1,\>s=p_2$ and the two odd fields $\rho=\zeta_1,\>\sigma=\zeta_2$.
Indeed the equations to be solved are:
\[
\left\{\begin{array}{l}
z^2 (p^2-p)+z(p_x-p_1p)=p_{1,x}+p_2;\\
z^2 (2p\zeta-\zeta)+z(\zeta_x-\zeta_1p^2-p_1\zeta)
=\zeta_1p+\zeta_2+p_1p\zeta,\\ 
q_1\big(z(p^2-p)+p_x-p_1p\big)=z(2pp_x-p_x)+p_{xx}-p_{1x}p-p_1p_x.
\end{array}\right.
\]
From the first one we get
\[
p_{k+2}=-p_{k+1\,x}-\sum_{j=1}^kp_j p_{k-j+2},\quad k\ge1
\]
and from the second a similar formula expressing $\zeta_{j+2},\> j\ge 1$ 
in terms of $\zeta_1,\zeta_2$ and the $p_k$'s.
Plugging the first equation into the third one, we finally get
\[
q_1=\frac{d}{dx}\log(p_2+p_{1x}).
\]
The resulting equations of motion relative to the time $t_4$ are the
following:
\begin{equation}
  \label{eq:sakns}
  \left\{\begin{split}
\dot{r}&=\big(r_x+2s-r^2\big)_x;\\
\dot{s}&=-\big(s_x+rs\big)_x;\\
\dot{\rho}&=\rho_{xx}+2\sigma_x-2r\rho_x-2r_x\rho+2\big(\log
(r_x+s)\big)_x\big(\rho_x+\sigma-r\rho\big)\\ 
\dot{\sigma}&=-\sigma_{xx}+2r\rho_{xx}-2(r^2+s)\rho_x-2r_x\sigma+2r\big(\log
(r_x+s)\big)_x\big(\rho_x+\sigma-r\rho\big).\end{split}\right.
\end{equation}
We notice that these evolution equations for $r$ and $s$ coincide 
with those of the realization of the well-known AKNS (or two--boson) 
\ger y as a rational reduction of the KP \ger y~\cite{Kr95,Rat}.
\eexample

\section{Linearization}\label{linear}

The evolution equations of SCS we have introduced in Section~\ref{SCS} 
are not linear, and not directly linearizable.
To obtain their ``linearized version'', allowing to provide explicit 
solutions, we can exploit Darboux covering techniques as it has been done 
in \cite{FMP} for KP.
The idea is to find a Darboux covering which intertwines the super Central
System SCS defined in Section \ref{FdBSKP} with a new hierarchy whose
linearization can be achieved by elementary methods.

To this end, let $\wid{\cal M}$ be the space of sequences of Laurent series
$\{\SY^{(k)}\}_{k\ge 0}$ of the form
$$\left\{\begin{array}{l}
\SY^{(2k)}=z^{k}+
\sum_{j>0}\left(\SY^{2k}_{0,j}z^{-j}+
\SY^{2k}_{1,j}\theta z^{-j}\right) \\
\SY^{(2k+1)}=\theta z^{k}+
\sum_{j>0}\left(\SY^{2k+1}_{0,j}z^{-j}+
\SY^{2k+1}_{1,j}\theta z^{-j}\right)\>,
\end{array}\right.$$
where $\overline{\SY^{(k)}}=k\bmod 2$.
The third manifold $\wid{\cal P}$ of Definition \ref{covering} is just a
copy of $\wid{\cal M}$ formed by the sequences $\{\SH{k}\}_{k\ge 0}$.
Finally, the manifold $\wid{\cal N}$ is the Cartesian product
$\wid{\cal M}\times\wid{\cal G}$ of $\wid{\cal M}$ by the group of even
invertible formal power series $\hat{w}$ of the form
$$\hat{w}=1+\sum_{j>0}\left(\hat{w}_{0,j}z^{-j}+
\hat{w}_{1,j}\theta z^{-j}\right).$$

The next step is to define suitable vector fields $\wid{\cal X}$,
$\wid{\cal Y}$ and $\wid{\cal Z}$ on $\wid{\cal M}$, $\wid{\cal N}$ and
$\wid{\cal P}$, respectively.
The vector field $\wid{\cal Z}$ is any vector field of SCS, which is
completely characterized by
$$(\de_{t_k}+\SH{k})\cdot W\subset W.$$
The flow can be identified by using an index, so we call this vector field
$\wid{\cal Z}_k$.
To define $\wid{\cal X}$ we introduce the subspace $W^{(\SY)}$ of $V$
spanned by the $\SY^{(j)}$'s.
Then, if $k=2n$ is even we let $\wid{\cal X}_k$ be the vector field
characterized by the property
$$(\de_{t_k}+z^{n})\cdot W^{(\SY)}\subset W^{(\SY)},$$
while if $k=2n+1$ we let $\wid{\cal X}_k$ be the vector field 
characterized by
$$(\de_{t_k}+\theta z^{n})\cdot W^{(\SY)}\subset W^{(\SY)}.$$
As for SCS, we can write down the equations defining $\wid{\cal X}_k$ by
comparing coefficients: if $k=2n$
$$\de_{t_k}\SY^{(j)}+z^n\SY^{(j)}=\SY^{(j+2n)}+
\sum_{l=1}^n\left(\SY^{j}_{0,l}\SY^{(2n-2l)}+
\SY^{j}_{1,l}\SY^{(2n-2l+1)}\right),$$
while if $k=2n+1$
$$\left\{\begin{array}{l}
\de_{t_k}\SY^{(2j)}+\theta z^n\SY^{(2j)}=\SY^{(2j+2n+1)}+
\sum_{l=1}^n\SY^{2j}_{0,l}\SY^{(2n-2l+1)} \\
\de_{t_k}\SY^{(2j+1)}+\theta z^n\SY^{(2j+1)}=
-\sum_{l=1}^n\SY^{2j+1}_{0,l}\SY^{(2n-2l+1)}\>.
\end{array}\right.
$$
The definition of $\wid{\cal Y}_k$ is obtained imposing the further condition
$$(\de_{t_k}+z^n)\cdot\hat{w}\in W^{(\SY)}\mbox{ if } k=2n,$$
$$(\de_{t_k}+\theta z^n)\cdot\hat{w}\in W^{(\SY)} \mbox{ if } k=2n+1\>.$$
As in \cite{FMP}, we give the following
\begindef
We call {\em super Sato System} (SS) the family of
vector fields $\{\wid{\cal X}_k\}_{k>0}$ on $\wid{\cal M}$ and
{\em super Darboux--Sato System} (SDS) the family of vector fields
$\{\wid{\cal Y}_k\}_{k>0}$ on $\wid{\cal N}$.
\enddef
The next step is to define the maps $\hat{\mu}:\wid{\cal N}\to\wid{\cal M}$
and $\hat{\sigma}:\wid{\cal N}\to\wid{\cal P}$.
The first is as usual the projection
$$(\{\SY^{(k)}\}_{k\ge 0},\hat{w})\mapsto\{\SY^{(k)}\}_{k\ge 0},$$
while the second is defined by imposing the intertwining condition
$$\hat{w}\cdot W=W^{(\SY)},$$
which holds if and only if
$$\left\{\begin{array}{l}
\hat{w}\SH{2j}=\SY^{(2j)}+
\sum_{l=1}^j\left(\hat{w}_{0,l}\SY^{(2j-2l)}+
\hat{w}_{1,l}\SY^{(2j-2l+1)}\right) \\
\hat{w}\SH{2j+1}=\SY^{(2j+1)}+
\sum_{l=1}^j\hat{w}_{0,l}\SY^{(2j-2l+1)}\>.
\end{array}\right.$$
\begin{lemma}
The SDS system is a Darboux intertwiner of SS with SCS.
\end{lemma}
\proof
We need only to prove that $\hat{\sigma}_*(SDS)=SCS$.
This follows by observing that the definitions of SDS and $\hat{\sigma}$ imply
$$\left\{\begin{array}{l}
\de_{t_{2k}}\hat{w}+z^k\hat{w}=\hat{w}\SH{2k} \\
\de_{t_{2k+1}}\hat{w}+\theta z^k\hat{w}=\SH{2k+1}\>,
\end{array}\right.$$
so
$$\left\{\begin{array}{l}
\hat{w}\cdot(\de_{t_{2k}}+\SH{2k})=(\de_{t_{2k}}+z^k)\cdot\hat{w} \\
\hat{w}\cdot(\de_{t_{2k+1}}+\SH{2k+1})=(\de_{t_{2k}}+\theta 
z^k)\cdot\hat{w}\>. \end{array}\right.$$
Hence, we get
\begin{eqnarray*}
\hat{w}\cdot(\de_{t_{2k}}+\SH{2k})\cdot W & = &
(\de_{t_{2k}}+z^k)\cdot\hat{w}W \\
& = & (\de_{t_{2k}}+z^k)\cdot W^{(\SY)}\subset W^{(\SY)}
\end{eqnarray*}
and
\begin{eqnarray*}
\hat{w}\cdot(\de_{t_{2k+1}}+\SH{2k+1})\cdot W & = &
(\de_{t_{2k+1}}+\theta z^k)\cdot\hat{w}W \\
& = & (\de_{t_{2k+1}}+\theta z^k)\cdot W^{(\SY)}\subset W^{(\SY)},
\end{eqnarray*}
showing that $(\de_{t_k}+\SH{k})\cdot W\subset W$, i.e. the SCS. \endpf
We consider now the map $\hat{\rho}:\wid{\cal M}\to\wid{\cal N}$ defined by
$$\{\SY^{(k)}\}_{k\ge 0}\mapsto(\{\SY^{(k)}\}_{k\ge 0},\SY^{(0)})$$
and the corresponding map
$\hat{\sigma}\circ\hat{\rho}:\wid{\cal M}\to\wid{\cal P}$.
\begin{lemma}
The submanifold $\hat{\rho}(\wid{\cal M})$ of $\wid{\cal N}$ is a section of
$\hat{\mu}$ invariant under SDS.
\end{lemma}
\proof
The previous definitions imply
$$\left\{\begin{array}{l}
\de_{2j}(\hat{w}-\SY^{(0)})=-z^j(\hat{w}-\SY^{(0)}) \\
\phantom{12345}+\sum_{l=1}^j\left((\hat{w}_{0,l}-\SY^0_{0,l})
\SY^{(2j-2l)}+
(\hat{w}_{1,l}-\SY^0_{1,l})\SY^{(2j-2l+1)}\right) \\
\\
\de_{2j+1}(\hat{w}-\SY^{(0)})=-\theta z^j(\hat{w}-\SY^{(0)})+
\sum_{l=1}^j(\hat{w}_{0,l}-\SY^0_{0,l}))\SY^{(2j-2l+1)}
\end{array}\right.,$$
proving the lemma.
\endpf
We have now to linearize the super Sato system.
To achieve the result it is better to introduce the infinite even matrix
$\WY$ defined by
$${\WY}_{jk}:=\left\{\begin{array}{ll}
\SY^j_{0,\frac{k+2}{2}} & \mbox{for $k$ even} \\
\\
\SY^j_{1,\frac{k+1}{2}} & \mbox{for $k$ odd},
\end{array}\right.$$
where $j,k\ge 0$, and the associated matrix $\widetilde{\WY}$ whose entries are
$$\widetilde{\WY}_{jk}:=(-1)^{\bar{\WY}_{jk}}{\WY}_{jk}=
(-1)^{j+k}{\WY}_{jk}.$$
An easy computation shows that the flows of the SS hierarchy translate
into the following Riccati type evolution equations:
\begin{equation}\label{MRE}
\left\{\begin{array}{l}
\de_{2n}{\WY}+{\WY}{\Lambda_2^t}^n-\Lambda_2^n{\WY}=
{\WY}\Gamma_{2n}{\WY} \\
\de_{2n+1}{\WY}+\widetilde{\WY}\Lambda_1{\Lambda_2^t}^n-
\Lambda_1\Lambda_2^n{\WY}=\widetilde{\WY}\Gamma_{2n+1}{\WY}
\end{array}\right.,
\end{equation}
where $^t$ means ordinary transposition (not super transposition), $\Lambda_1$
is the odd shift matrix with entries
$$(\Lambda_1)_{jk}:=\frac{1-(-1)^k}{2}\delta_{k,j+1},$$
$\Lambda_2$ is the even shift matrix with entries
$$(\Lambda_2)_{jk}:=\delta_{k,j+2},$$
$\Gamma_{2n}$ is the even convolution matrix defined by
$$(\Gamma_{2n})_{jk}:=\frac{1-(-1)^k}{2}\delta_{k,2n-j}+
\frac{1-(-1)^{k+1}}{2}\delta_{k,2n-j-2}$$
and finally $\Gamma_{2n+1}$ is the odd convolution matrix given by
$$(\Gamma_{2n+1})_{jk}:=\frac{1-(-1)^k}{2}\delta_{k,2n-j-1}.$$
Observe that these matrices satisfy the relations
$$[\Lambda_1,\Lambda_1]=[\Lambda_1,\Lambda_2]=[\Lambda_1,\Lambda_2^t]=0,$$
$$\Lambda_2^t\Gamma_n=\Gamma_n\Lambda_2,\>
\Lambda_1\Gamma_{2n}=\Gamma_{2n}\Lambda_1,\> \mbox{ and }\>
\Lambda_1\Gamma_{2n+1}=\Gamma_{2n+1}\Lambda_1=0,$$
which imply the compatibility of the above system of matrix Riccati equations.
\begin{proposition}
The infinite even matrix $\WY$ is a solution of \rref{MRE} if and only 
if it has the form
${\WY}={\VW}\cdot{\UW}^{-1}$, where $\UW$ and $\VW$ are
infinite even matrices satisfying the constant coefficients linear system
$$\left\{\begin{array}{l}
\de_{2n}{\UW}={\Lambda_2^t}^n{\UW}-\Gamma_{2n}{\VW} \\
\de_{2n+1}{\UW}=\Lambda_1{\Lambda_2^t}^n{\UW}-\Gamma_{2n+1}{\VW} \\
\de_{2n}{\VW}=\Lambda_2^n{\VW} \\
\de_{2n+1}{\VW}=\Lambda_1\Lambda_2^{n}{\VW}
\end{array}\right.$$
with, of course, $\UW$ invertible.
\end{proposition}
\proof
The proof is exactly the same as in the commutative case, once we have observed
that for two matrices $\UW$ and $\VW$ the following relations hold:
$$\left\{\begin{array}{l}
\de_{t_{2k}}({\UW}{\VW})=(\de_{t_{2k}}{\UW}){\VW}+
{\UW}(\de_{t_{2k}}{\VW}) \\
\de_{t_{2k+1}}({\UW}{\VW})=(\de_{t_{2k+1}}{\UW}){\VW}+
\widetilde{\UW}(\de_{t_{2k+1}}{\VW}) \\
\widetilde{\UW\VW}=\widetilde{\UW}\widetilde{\VW}\Rightarrow
\widetilde{{\UW}^{-1}}=\widetilde{\UW}^{-1}
\end{array}\right..$$
Thus, if $\UW$ and $\VW$ solve the system of linear equations of the
statement and if we let ${\WY}={\VW}{\UW}^{-1}$, then
\begin{eqnarray*}
\de_{2n}{\WY} & = & (\de_{2n}{\VW}){\UW}^{-1}-
{\VW}{\UW}^{-1}(\de_{2n}{\UW}){\UW}^{-1} \\
& = & \Lambda_2^n{\VW\UW}^{-1}-{\VW\UW}^{-1}{\Lambda_2^t}^n+
{\VW\UW}^{-1}\Gamma_{2n}{\VW\UW}^{-1} \\
& = & -{\WY}{\Lambda_2^t}^n+\Lambda_2^n{\WY}+{\WY}\Gamma_{2n}{\WY}
\end{eqnarray*}
and
\begin{eqnarray*}
\de_{2n+1}{\WY} & = & (\de_{2n+1}{\VW}){\UW}^{-1}-
\widetilde{\VW}\widetilde{\UW}^{-1}(\de_{2n+1}{\UW}){\UW}^{-1} \\
& = & \Lambda_1\Lambda_2^n{\VW\UW}^{-1}-
\widetilde{\VW}\widetilde{\UW}^{-1}\Lambda_1{\Lambda_2^t}^n+
\widetilde{\VW}\widetilde{\UW}^{-1}\Gamma_{2n+1}{\VW\UW}^{-1} \\
& = & -\widetilde{\WY}\Lambda_1{\Lambda_2^t}^n+
\Lambda_1\Lambda_2^n{\WY}+\widetilde{\WY}\Gamma_{2n}{\WY}.
\end{eqnarray*}
Therefore, if we look for a solution $\WY$ of the Riccati matrix equations
of SS with initial condition ${\WY}(0)={\WY}_0$, we have simply to solve
the linear system above imposing the initial conditions
${\VW}(0)={\WY}_0$ and ${\UW}(0)={\mathbb I}$.
As we already noticed, the necessary condition
$$\Gamma_{2n+1}\Lambda_1=0$$
for the integrability of the linear system holds.
\endpf
Of course, the computations given in the proposition are only formal: to
make sense of them one should also introduce a suitable notion of convergence
for the intervening series in infinite variables.
However, notice that the constraint ``${\WY}_{jk}=0$ when either $j\ge J$ or
$k\ge K$'' is compatible with the evolution equations for $\WY$, allowing us
to consider reductions where only the finite submatrix ${\WY}_{JK}$ of
$\WY$ consisting of its first $J$ rows and $K$ columns does not vanish.
Obviously, ${\WY}_{JK}$ evolves according to the reduced Riccati equations
$$\left\{\begin{array}{l}
\de_{2n}{\WY}_{JK}+
{\WY}_{JK}{\Lambda_{2,KK}^t}^n-\Lambda_{2,JJ}^n{\WY}_{JK}=
{\WY}_{JK}\Gamma_{2n,KJ}{\WY}_{JK} \\
\de_{2n+1}{\WY}_{JK}+
\widetilde{\WY}_{JK}(\Lambda_1{\Lambda_2^t}^n)_{KK}-
\Lambda_{1,JJ}\Lambda_{2,JJ}^n{\WY}_{JK}=
\widetilde{\WY}_{JK}\Gamma_{2n+1,KJ}{\WY}_{JK} \\
\end{array}\right. .$$
This is a closed system of (graded) ordinary differential equations in a
finite number of variables. It
yields ``finite type'' solutions (i.e. depending only on finitely many
times) of SS and hence of SCS and HSKP.
Observe that the compatibility of the reduced system requires $K$ to be even;
in this case
$(\Lambda_1{\Lambda_2^t}^n)_{KK}=\Lambda_{1,KK}{\Lambda_{2,KK}^t}^n$.
\subsection{An explicit example}
To show an example, we compute the solution of SS associated to $J=3$
and $K=4$.
To simplify notations let us call $Y:={\WY}_{34}$,
$$A_1:=\Lambda_{1,33}=\left(\begin{array}{ccc}
0 & 1 & 0 \\
0 & 0 & 0 \\
0 & 0 & 0
\end{array}\right),\quad
A_2:=\Lambda_{2,33}=\left(\begin{array}{ccc}
0 & 0 & 1 \\
0 & 0 & 0 \\
0 & 0 & 0
\end{array}\right),$$
$$B_1:=\Lambda_{1,44}=\left(\begin{array}{cccc}
0 & 1 & 0 & 0 \\
0 & 0 & 0 & 0 \\
0 & 0 & 0 & 1 \\
0 & 0 & 0 & 0
\end{array}\right),\quad
B_2:=\Lambda_{2,44}^t=\left(\begin{array}{cccc}
0 & 0 & 0 & 0 \\
0 & 0 & 0 & 0 \\
1 & 0 & 0 & 0 \\
0 & 1 & 0 & 0
\end{array}\right),$$
and $C_k:=\Gamma_{k,43}$.
The relevant (i.e. different from zero) convolution matrices are
$$C_2:=\left(\begin{array}{ccc}
1 & 0 & 0 \\
0 & 1 & 0 \\
0 & 0 & 0 \\
0 & 0 & 0
\end{array}\right),\quad
C_3:=\left(\begin{array}{ccc}
0 & 1 & 0 \\
0 & 0 & 0 \\
0 & 0 & 0 \\
0 & 0 & 0
\end{array}\right),$$
$$C_4:=\left(\begin{array}{ccc}
0 & 0 & 1 \\
0 & 0 & 0 \\
1 & 0 & 0 \\
0 & 1 & 0 \\
\end{array}\right),\quad
C_5:=\left(\begin{array}{ccc}
0 & 0 & 0 \\
0 & 0 & 0 \\
0 & 1 & 0 \\
0 & 0 & 0
\end{array}\right),\quad
C_6:=\left(\begin{array}{ccc}
0 & 0 & 0 \\
0 & 0 & 0 \\
0 & 0 & 1 \\
0 & 0 & 0
\end{array}\right).$$
We see that $A_2^2=0$ and $B_2^2=0$, so the solution of SS (or SCS) will depend
only on the first six times.
We solve the Riccati system for $Y$ by introducing the $4\times 4$ matrix $U$
and the $3\times 4$ matrix $V$ which are solutions of the following linear
Cauchy problems:
$$\left\{\begin{array}{l}
\de_{t_{2k}}V=A_2^kV \\
\de_{t_{2k+1}}V=A_1A_2^kV \\
\\
V(0)=Y(0)
\end{array}\right.\qquad
\left\{\begin{array}{l}
\de_{t_{2k}}U=B_2^kU-C_{{2k}}V \\
\de_{t_{2k+1}}U=B_1B_2^{k}U-C_{{2k+1}}V \\
\\
U(0)={\mathbb I}
\end{array}\right.$$
and then putting $Y:=VU^{-1}$.
First of all we find that
$$V=\exp\sum_{j>0}(t_{2j}A_2^j+t_{2j-1}A_1A_2^{j-1})V(0)=
\left(\begin{array}{ccc}
1 & t_1 & t_2 \\
0 & 1 & 0 \\
0 & 0 & 1
\end{array}\right)Y(0).$$
Then we solve the system for $U$ by introducing the matrix $U_0$ defined by
$$U=\exp\sum_{j>0}(t_{2j}B_2^j+t_{2j-1}B_1B_2^{j-1})U_0=
({\mathbb I}+t_1B_1+t_2B_2+(t_3+t_1t_2)B_1B_2)U_0$$
and evolving as
$$\left\{\begin{array}{l}
\de_{t_{2k}}U_0=-({\mathbb I}-t_1B_1-t_2B_2-(t_3-t_1t_2)B_1B_2)C_{{2k}}V \\
\de_{t_{2k+1}}U_0=-({\mathbb I}+t_1B_1-t_2B_2+(t_3-t_1t_2)B_1B_2)C_{2k+1}V
\end{array}\right..$$
The equations for $U_0$ are easily solvable and we get
$$U_0={\mathbb I}-\left(\begin{array}{ccc}
t_2 & t_3 & t_4+\frac{1}{2}t_2^2 \\
0 & t_2 & 0 \\
t_4-\frac{1}{2}t_2^2 & t_5-t_2t_3 & t_6-\frac{1}{3}t_2^3 \\
0 & t_4-\frac{1}{2}t_2^2 & 0
\end{array}\right)Y(0).$$
In order to write down an effective solution, we choose simple initial
conditions, e.g.
$$Y(0)=\left(\begin{array}{cccc}
0 & 0 & 0 & 0 \\
0 & 0 & 0 & 0 \\
0 & 0 & 1 & 0
\end{array}\right).$$
Then
$$V=\left(\begin{array}{cccc}
0 & 0 & t_2 & 0 \\
0 & 0 & 0 & 0 \\
0 & 0 & 1 & 0
\end{array}\right),$$
$$U=\left(\begin{array}{cccc}
1 & t_1 & -t_4-\frac{1}{2}t_2^2 & 0 \\
0 & 1 & 0 & 0 \\
t_2 & t_3+t_1t_2 & 1-t_6-t_2t_4-\frac{1}{6}t_2^3 & t_1 \\
0 & t_2 & 0 & 1
\end{array}\right).$$
Finally, we find
$$\SY^{(0)}=1-\frac{3t_2^2}{\tau}z^{-1}+\frac{3t_2(t_1t_2-t_3)}{\tau}\theta 
z^{-1} +\frac{3t_2}{\tau}z^{-2}-\frac{3t_1t_2}{\tau}\theta z^{-2},$$
$$\SY^{(1)}=\theta,$$
$$\SY^{(2)}=z-\frac{3t_2}{\tau}z^{-1}+\frac{3(t_1t_2-t_3)}{\tau}\theta 
z^{-1} +\frac{3}{\tau}z^{-2}-\frac{3t_1}{\tau}\theta z^{-2},$$
$$\SY^{(2k)}=z^k\qquad\mbox{for $k>1$}$$
$$\SY^{(2k+1)}=\theta z^k\qquad\mbox{for $k>0$},$$
where $\tau=3+t_2^3-3t_6$.
We can thus compute the first super currents of SCS
$$\SH{1}=
\theta+\sum_{k>0}\left(\frac{3t_2^2}{\tau}z^{-1}+\frac{3t_2}
{\tau}z^{-2}\right)^k\theta,$$
\begin{eqnarray*}
\SH{2} & = & z-\frac{3t_2^2}{\tau}+3\sum_{k\ge 0}
\left(\frac{3t_2^2}{\tau}z^{-1}-\frac{3t_2}{\tau}z^{-2}\right)^k\times \\
& &
\left(\frac{t_2^2}{\tau}-\frac{2t_2}{\tau}z^{-1}+\frac{2t_1t_2-t_3}{\tau}
\theta z^{-1}
+\frac{1}{\tau}z^{-2}-\frac{t_1}{\tau}\theta z^{-2}\right).
\end{eqnarray*}
As explained in Section \ref{SCS}, we obtain a solution of HSKP after
substituting $t_2$ and $t_1$ with $x$ and $\varphi+t_1$ respectively and
putting $\hat{h}=\SH{1}+\varphi\SH{2}$:
$$\left\{\begin{array}{l}
\nu=0 \\
a=1+\sum_{k>0}\left(\frac{3x^2}{\tau}z^{-1}-\frac{3x}{\tau}z^{-2}\right)^k \\
h=z-\frac{6x}{\tau}z^{-1}+\frac{3}{\tau}z^{-2}+
3\sum_{k>0}\left(\frac{3x^2}{\tau}z^{-1}-\frac{3x}{\tau}z^{-2}\right)^k
\left(\frac{x^2}{\tau}-\frac{2x}{\tau}z^{-1}+\frac{1}{\tau}z^{-2}\right) \\
\psi=3\sum_{k\ge 
0}\left(\frac{3x^2}{\tau}z^{-1}-\frac{3x}{\tau}z^{-2}\right)^k
\left(\frac{2t_1x-t_3}{\tau}z^{-1}-\frac{t_1}{\tau}z^{-2}\right)\>.
\end{array}\right.$$
\subsection*{Acknowledgments} We would like to thank F. Magri 
 for useful comments and suggestions. 
This work was partially supported by the GNFM branch 
of the {\em Istituto Nazionale di Alta Matematica}, and by the Italian MURST, 
 under the Cofin99 project {\em Geometry of Integrable Systems}.

\end{document}